\documentclass{aa}  
\usepackage{graphicx}

\usepackage[colorlinks=true, linkcolor=blue, citecolor=blue, urlcolor=blue]{hyperref}

\usepackage{txfonts}
\usepackage{xcolor}
\usepackage{array}
\usepackage{float}
\usepackage{placeins} 
\usepackage{caption}

\newcommand{\oiii}{[O\,{\sc iii}]$\lambda$5007}
\newcommand{\sii}{[S\,{\sc ii}]$\lambda$6716, 6731}
\newcommand{\nii}{[N\,{\sc ii}]$\lambda$6548,6583}
\newcommand{\kms}{\,km\,s$^{-1}$}
\newcommand{\ergs}{\,erg\,s$^{-1}$}

\newcommand{\weighty}{W$_{80}$}
\newcommand{\vten}{v$_{10}$}

\begin{document}

   \title{Connecting outflows with radio emission in active galactic nuclei at cosmic noon}
\titlerunning{Connecting outflows with radio emission in AGNs at cosmic noon}

   \author{Gabriele S. Ilha
          \inst{1,2},
          C. M. Harrison
          \inst{3},
          V. Mainieri\inst{1},
          Ann Njeri\inst{3},
          E. Bertola\inst{4},
          M. Bischetti\inst{5,6},
          C. Circosta\inst{7,8},
          C. Cicone\inst{18},
          G. Cresci\inst{4},
          V. A. Fawcett\inst{3}, 
          A. Georgakakis\inst{9}, 
          D. Kakkad\inst{10,11}, 
          I. Lamperti\inst{4,12}, 
          A. Marconi\inst{4,12}, 
          M. Perna\inst{13},
          A. Puglisi\inst{14}, 
          D. Rosario\inst{3}, 
          G. Tozzi\inst{15}, 
          C.Vignali\inst{16,17}
          G. Zamorani\inst{16}
          }
\authorrunning{Gabriele S. Ilha et al.} 
   \institute{European Southern Observatory, Karl-Schwarzschild-Str. 2, 85748 Garching bei München, Germany\\
        \email{ gabriele.ilha@acad.ufsm.br }
         \and Universidade de São Paulo, Instituto de Astronomia, Geofísica e Ciências Atmosféricas, São Paulo, SP, Brazil
        \and School of Mathematics, Statistics and Physics, Newcastle University, Newcastle upon Tyne, NE1 7RU, UK
        \and INAF-Osservatorio Astrofisico di Arcetri, Largo E. Fermi 5, I-50125 Florence, Italy
        \and Dipartimento di Fisica, Universitá di Trieste, Sezione di Astronomia, Via G.B. Tiepolo 11, I-34131 Trieste, Italy
        \and  INAF—Osservatorio Astronomico di Trieste, Via G. B. Tiepolo 11, I–34131 Trieste, Italy
        \and ESA, European Space Astronomy Centre (ESAC), Camino Bajo del Castillo s/n, 28692 Villanueva de la Cañada, Madrid, Spain
        \and Institut de Radioastronomie Millimétrique (IRAM), 300 rue de la Piscine, 38400 Saint-Martin-d’Hères, France
        \and Institute for Astronomy \& Astrophysics, National Observatory of Athens, V. Paulou \& I. Metaxa, 11532, Greece
        \and Centre for Astrophysics Research, Department of Physics, Astronomy and Mathematics, University of Hertfordshire, Hatfield, AL10 9AB, UK
        \and Space Telescope Science Institute, 3700 San Martin Drive, Baltimore, 21218,USA
        \and Dipartimento di Fisica e Astronomia, Università di Firenze, Via G. Sansone 1, I-50019, Sesto F.no (Firenze), Italy
        \and Centro de Astrobiología (CAB), CSIC–INTA, Cra. de Ajalvir Km. 4, 28850 – Torrejón de Ardoz, Madrid, Spain
        \and School of Physics and Astronomy, University of Southampton, Highfield SO17 1BJ, UK
        \and Max-Planck-Institut für Extraterrestrische Physik (MPE), Giessenbachstraße 1, D-85748, Garching, Germany
        \and INAF- Osservatorio di Astrofisica e Scienza dello Spazio di Bologna, via Piero Gobetti, 93/3, I-40129 Bologna, Italy
        \and
        Dipartimento di Fisica e Astronomia ‘Augusto Righi’, Universitá degli Studi di Bologna, via P. Gobetti, 93/2, 40129 Bologna, Italy
        \and Institute of Theoretical Astrophysics, University of Oslo, PO Box 1029, Blindern, 0315, Oslo, Norway
             }

   \date{Received May XX, 2025; accepted May XX, 2025}

  \abstract
   {Active galactic nucleus (AGN) feedback is a well-known mechanism in the evolution of galaxies. However, constraining its parameters remains a significant challenge. One open question is the driving mechanism of galaxy-scale outflows. At low redshift, radio jets often interact with the interstellar medium (ISM), generating turbulence and driving ionized outflows.}
   {Despite this evidence at low redshift, relatively few studies have investigated the radio-ionized gas connection at cosmic noon. Thus, our main goal is to conduct a pilot study using Very Large Array (VLA) data for three quasars ($z\sim$2.0) with moderate to high radio power ($\sim 10^{24.86}-10^{28.15}$ W Hz$^{-1}$) that have ionized outflows identified in observations from the SUPER Survey to investigate whether this connection also exists. }
   {We used \oiii\ data  from VLT/SINFONI analyzed in earlier studies along with new 6.2 GHz VLA radio observations at comparable spatial resolution ($\sim$0.3\arcsec\---0.5\arcsec\ or  2.5--4.2 kpc). We also incorporated radio data from the literature at different frequencies and resolutions to explore the radio emission.}
   {We detected an extended radio structure in our VLA A-array data for two quasars: J1333+1649 and CID-346. The extended structure in J1333+1649 ($\sim$0.5\arcsec\ or 4.16 kpc) aligns with the smaller-scale emission ($\sim$0.01\arcsec\ -- 0.02\arcsec\ or 0.08--0.17 kpc) seen in archival images, suggesting a jet propagating from nuclear to galaxy-wide scales. In all three quasars, we found that the brightest radio emission and ionized gas have comparable spatial scales. Furthermore, the position angles of the radio emission and ionized gas present offsets smaller than 30$^{\circ}$ for the two targets with extended structures. Given that the kinematics of the ionized gas in all three quasars is dominated by outflows, our results suggest a strong connection between radio emission and ionized outflows in typical AGNs at cosmic noon.}
   { This result  is similar to what has been previously observed in radio-powerful AGN at the same epoch and in  AGN at lower redshifts. Based on energetic considerations and comparisons with archival data, radio jets could be a significant mechanism for driving outflows in AGN from cosmic noon to low redshifts. However, with the exception of one object (J1333+1649), we cannot rule out the possibility that the radio emission arises from shocks in the ISM caused by disk winds or radiatively driven outflows. Further studies on larger samples are required to determine whether radio jets are driving the observed outflows.
 }
   \maketitle

\section{Introduction}
Active galactic nucleus (AGN) feedback plays a crucial role in our understanding of galaxy evolution. AGNs are thought to regulate or suppress star formation in their host galaxies \citep{Carniani+16,Wylezalek+16,Mercedes-Feliz+23} by injecting energy into the surrounding  medium or expelling it through outflows or jets. Thus, AGN feedback emerges as a key mechanism regulating galaxy growth \citep{DiMatteo+05,Springel+05,Choi+18,dave+19,van+19}. Following the terminology derived from the earliest semi-analytical models and hydrodynamic simulations  \citep[e.g.,][]{DiMatteo+05,Springel+05,Sijacki+07,Dubois+10}, feedback operates in two primary modes: the radio (or kinetic) mode and the quasar (or radiative) mode. The radio mode is associated with AGNs that exhibit low mass accretion rates and low luminosities, and it is commonly linked to the effects of radio jets. In contrast, the quasar mode involves radiative energy being injected into the surrounding gas through radiatively driven winds. This typically occurs in AGNs accreting at high mass accretion rates, with luminosities near the Eddington limit \citep[see reviews][]{Fabian+12,Harrison+24}. Although this terminology is still used in the literature, it is now well established that radio emission associated with AGN occurs in both high and low accretion rate regimes, and may trace feedback-related processes even in moderate to low radio luminosity sources \citep[e.g.,][]{Zakamska+14,Kellermann+16,Jarvis+19,Macfarlane+21,Roy+24,Harrison+24,Fawcett+25,Njeri+25}. In other words, there are quasars whose radio emission from the AGN may be considered relevant for feedback. Thus, the most important feedback mechanisms for high accretion rate systems have not yet been well established.

One way to search for the impact of AGNs on their host galaxies is to look for outflows or highly disturbed gas in the interstellar medium (ISM). In this paper, we follow \citet{Harrison+24} and refer to ``outflows'' as the ISM gas that is carried by various driving mechanisms and use ``winds'' to describe the launching mechanism associated with accretion disks. Outflows are observed in different gas phases, such as ionized \citep{Crenshaw+10,Fiore+17,Kakkad+20,Riffel+23,Falcone+24,Tozzi+24,Liu+24,Gatto+24}, neutral \citep{Rupke+05,Morganti+05,Allison+15,Perna+19, Baron+21,Davies+24}, and molecular gas \citep{Cicone+12,Fiore+17,Fluetsch+18,Burillo+21,Ramos-Almeida+22,Lamperti+22,Riffel+23,Zhong+24,Bianchin+24,Costa-Souza+24,Holden+24}. AGN outflows have been identified in different gas phases by analyzing the kinematics, and we can explore their driving mechanism. In radiatively efficient AGNs (such as quasars), we can have different types of mechanisms driving the outflows, including radiation pressure, winds associated with the accretion disks, and low power collimated jets that could couple to the gas. In fact, several studies have explored the interplay between ionized gas and radio emission in quasars \citep{Zakamska+14,Hwang+18,Liao+19,Girdhar+22,Ulivi+24} and in low-luminosity AGNs \citep{Riffel+14,Woo+16,Venturi+21,Peralta+23,Hermosa+24}. Thus, in high accretion rate systems (such as quasars), the effect of radio emission has historically been underestimated in feedback studies. It may act as a key mechanism driving outflows and consequently play a more significant role in AGN feedback than initially considered. 

The connection between radio emission and outflows in radiatively efficient AGNs with moderate radio power has been observed in statistical samples with limited or no spatial information \citep[e.g.,][]{Mullaney+13,Zakamska+14,Villar+14,Villar+21,Alban+24,Kukreti+25,Escott+25} and in detailed spatially resolved studies \citep[ e.g.,][]{Venturi+21,Girdhar+22,Morganti+23,Ulivi+24,Girdhar+24,Speranza+24}. From the statistical spatially unresolved perspective, \citet{Mullaney+13} analyzed a sample of Sloan Digital Sky Survey (SDSS) AGNs at $z < 0.4$ with \oiii\ luminosity of $\sim 10^{40}-10^{45}$ \ergs\ and observed a correlation between disturbed gas kinematics and radio emission. According to \citet{Mullaney+13}, the most disturbed gas kinematics (associated with outflows) are found in radio-moderate AGNs. Similarly, \citet{Zakamska+14}, in their study of 568 quasars at $z < 0.8$ from SDSS, observed that outflow velocities in radio-quiet quasars correlate with radio luminosity, suggesting that relativistic particles are accelerated in shocks driven by ionized outflows. This finding aligns with the results of \citet{Alban+24,Kukreti+25}, who used large samples of AGNs with integral field spectroscopy at low redshift and report that radio-selected AGNs exhibit increased ionized gas line widths, which may be further boosted when the AGN have higher radiative luminosities (particularly in the central regions). 

\citet{Molyneux+19}, who also used an SDSS sample, found that extreme ionized outflows, characterized by components with a full width at half maximum (FWHM) greater than $ 1000$ \kms, are more likely to be associated with compact radio sources. The main result of \citet{Molyneux+19} has been analyzed more thoroughly using detailed spatially resolved data by \citet{Jarvis+19}, who investigated a sample of ten type-2 AGNs  at $z < 0.2$ classified as radio-quiet using high-resolution radio imaging and spatially resolved \oiii\ ionized gas kinematics that revealed outflows. They found that radio jets are the primary source of radio emission for most objects in the sample \citep[also see][] {Silpa+22, Njeri+25}. The radio jets were associated with increased turbulence and outflowing gas, demonstrating an interaction between the ionized gas and the jets. Based on these findings, they suggested that compact radio jets could interact with the ISM and drive outflows in radio-quiet AGNs.  Indeed, several other studies have suggested that compact, moderate power radio jets may have an important role in AGN feedback \citep[e.g.,][]{Villar+14, Villar+21,Venturi+21,Girdhar+22,Ramos-Almeida+22,Ayubinia+23,Audibert+23,Morganti+23,Girdhar+24,Roy+24,Audibert+25}.

Most of these studies were conducted at low redshift, but the peak of black hole growth (and hence the space density of radiatively efficient AGNs) and star formation occurs at cosmic noon \citep[$z\sim2$;][]{Madau+14}. However, there are relatively few studies exploring the connection between radio emission and outflows in this redshift range. \citet{Nesvadba+17} analyzed a sample of 33 powerful (radio power larger than 10$^{26}$ W Hz$^{-1}$) radio galaxies with $z$ > 2 using data from the Spectrograph for INtegral Field Observations in the Near Infrared (SINFONI) and the Very Large Array (VLA). Their study revealed several correlations between the properties of the ionized gas and the radio jets, including a clear alignment between the jet and gas axes as well as a strong correlation between the gas kinetic energy and the power of the radio source. In a single-object study, \citet{Cresci+23}, studying a quasar at $z=1.59$, found that a low-luminosity radio jet produces an expanding bubble from which an extended outflow emerges. 

One of our primary objectives is to investigate whether this  connection between radio jets and ionized outflows, potentially driven by compact jets, occurs in typical AGNs at cosmic noon ($z \sim 1-3$) using targets selected from the SUPER Survey \citep[SINFONI Survey for Unveiling the Physics and Effects of Radiative Feedback;][]{Circosta+18}. Therefore, we present a pilot study (first three targets) on a well-defined sample of radiatively efficient AGN with ionized gas winds from this  survey. Two out of the three objects exhibit a lower radio power than the targets of \citet{Nesvadba+17} (see Section \ref{sec:sample}). In this sense, our sample is complementary to their sample.

This paper is organized as follows: Section \ref{sec:sample}  introduces our sample. In Section \ref{sec:radio}, we describe the radio observations and ionized gas data utilized in this study. In Section \ref{sec:radio_results}, we describe the radio analysis and results. The connection between radio emission and ionized gas is presented in Section \ref{sec:radio_gas}. In Section \ref{sec:Radio_outflow}, we investigate the connection between radio emission and outflows in ionized gas, contextualizing our findings within previous results from the literature. Section \ref{sec:conclusions} provides a summary of our conclusions. Throughout this work, we adopted the following cosmological parameters: $\Omega_{M}$=0.3, $\Omega_{\Lambda}$ = 0.7, and $H_{0}$ = 70 \kms. All maps are shown with north up and east to the left.

\section{Sample}
\label{sec:sample}

\subsection{SUPER sample}
The targets were selected from the SUPER Survey \citep{Circosta+18,Mainieri+21}, a program designed to observe a sample of AGNs using the SINFONI instrument at the Very Large Telescope. A comprehensive description of the survey and sample is provided in \citet{Circosta+18}, while details of the observations and data reduction can be found in \citet{Kakkad+20} for type-1 and \citet{Tozzi+24} for  type-2 AGN. Here, we present only a brief overview of the key aspects. The primary objectives of this survey are  to investigate the presence of outflows in AGN host galaxies and their impact on star formation, using a sample selected independently of any prior assumptions about AGN feedback effects.

\begin{figure*}[htbp]
	\includegraphics[width=0.33\textwidth]{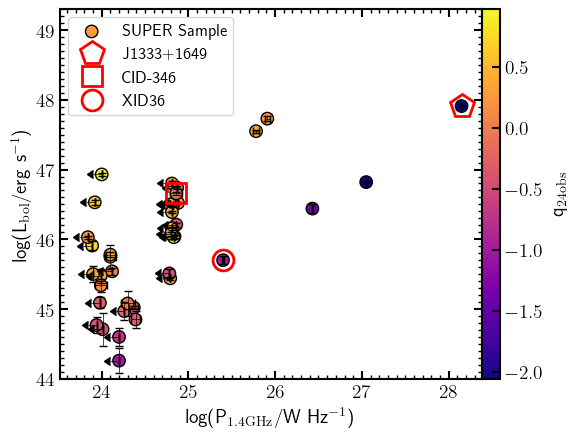}
        \includegraphics[width=0.33\textwidth]{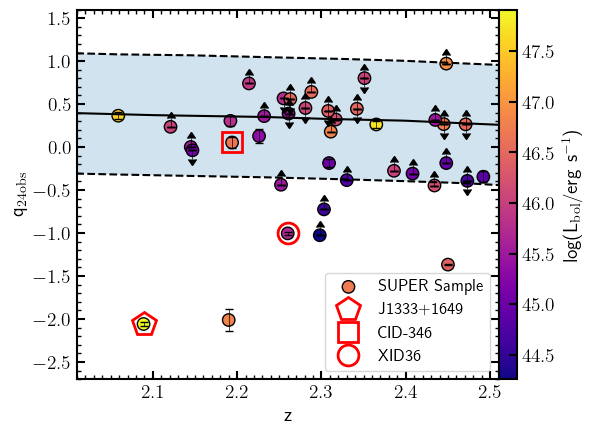}
        \includegraphics[width=0.33\textwidth]{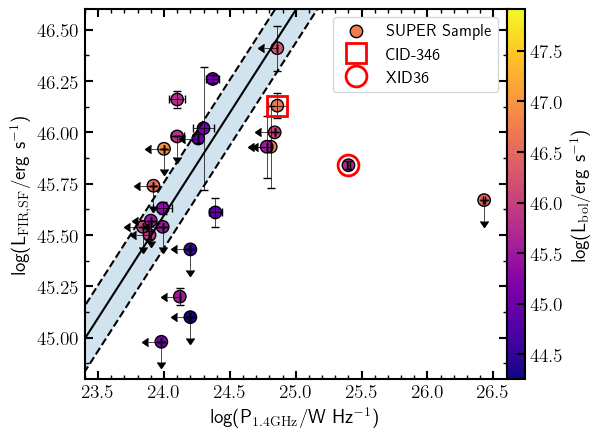}
    \caption{Properties of the SUPER sample highlighting the three targets of this work. Left: Bolometric luminosity (L$_{bol}$) vs.  radio power at 1.4 GHz.  The color bar shows the q$_{24obs}$ parameter, where  q$_{24obs}$ = log(S$_{24\mu\text{m}}$/S$_{r}$ ). S$_{24\mu\text{m}}$ and S$_{r}$ are the observed 24 $\mu\text{m}$ and 1.4 GHz flux densities. Middle: q$_{24obs}$ parameter vs. redshift.  The color bar shows the bolometric luminosity. The solid line represents the typical values for a star-forming galaxy \citep[M32; from][]{Bonzini+13}, while the dashed lines indicate the $\pm 2\sigma$ dispersion, which marks the locus of star-forming galaxies. Right: FIR luminosities vs. radio power at 1.4 GHz. The solid line is from \citet{Magnelli+15} and the dashed lines represent 2$\sigma$ relative to the solid line.  The color bar shows the bolometric luminosity.}
    \label{fig:sample}
\end{figure*}

The SUPER sample comprises 39 AGN with redshifts in the range of $2.0-2.5$. These AGN were selected based on their X-ray luminosity, $L_{X} > 10^{42}$\ergs, ensuring a robust AGN selection, including both type-1 and type-2 AGN \citep[see][]{Circosta+18,Kakkad+20,Vietri+20}. The sample includes 23 (58\%) type-1 and 16 (42\%) type-2 AGN \citep{Kakkad+20,Tozzi+24}. The X-ray luminosities obtained by \citet{Circosta+18} range from $10^{43}$ to $10^{46}$ \ergs. Additionally, \citet{Circosta+18} estimated the stellar masses ($\sim 4 \times 10^{9} - 2 \times 10^{11}$ M${\odot}$), star formation rates ($25 - 680$ M$_{\odot}$ yr$^{-1}$), and AGN bolometric luminosities ranging from $10^{44}$ to $10^{48}$ \ergs\ (as shown in Figure \ref{fig:sample}) via spectral energy distribution (SED) fitting of UV-to-far-infrared (FIR) photometry. This SED fitting of SUPER AGN was updated in \citet{Bertola+24}. Star formation rate (SFR) were constrained only for a subset of the sample with high-quality FIR data by \citet{Circosta+18}; as a result, we do not have  a SFR value for J1333+1649 (one of our targets). The majority of the sample presents stellar masses greater than 10$^{10}$ M$_{\odot}$ and log(SFR[M$_{\odot}$ yr$^{-1}$]) > 1.0 in the redshift range 2.0-2.5, thus making it representative of radiatively efficient AGNs at cosmic noon based on their stellar masses \citep[see][]{Aird+18} and SFRs \citep{Stanley+15,Aird+19}. Figure \ref{fig:sample}  presents the radio powers at 1.4 GHz and q$_{24 obs}$ = log(S$_{24\mu m}$/S$_{r}$) parameter from \citet{Circosta+18},  where S$_{24\mu m}$ are the observed flux densities at 24 $\mu$m  and S$_{r}$ the observed flux densities at 1.4 GHz. This parameter is commonly used as a measure of radio excess. The radio power and q$_{24 obs}$ values for the SUPER sample are in the range $\sim 10^{23.8}-10^{28.2}$ W Hz$^{-1}$ and approximately -2.0--1.0, as illustrated in Figure \ref{fig:sample}. In Section \ref{our_sample}, we discuss the properties of our sample.

The SINFONI observations were conducted as part of European Southern Observatory large program ($\sim$300 hours) 196.A-0377 (PI: V. Mainieri) between November 2015 and December 2018. These observations covered the H-band (1.45–1.85 $\mu$m) and K-band (1.95–2.45 $\mu$m) and were carried out in adaptive optics mode. The two bands encompass the rest-frame optical lines H$\beta$, \oiii, \nii, H$\alpha$, and \sii, which are key emission lines used to study AGN. These data have been used to determine the incidence of \oiii\ ionized outflows, and their physical properties, such as velocity, mass outflow rate, kinetic power \citep[see][]{Kakkad+20,Tozzi+24} and relationship to host galaxy star formation \citep[see][]{Lamperti+21, Kakkad+23}. The results in \citet{Kakkad+20} suggest that the radiative output of the AGN is capable of driving the observed outflows. However, a direct connection between the radio and outflow properties of the sample has yet to be investigated. This is important, especially given the incidence of jet-driven multiphase outflows seen even in the most bolometrically luminous AGN at low redshift \citep[e.g.,][]{Husemann+19,Jarvis+21,Girdhar+22,Audibert+23,Girdhar+24,Liao+24,Audibert+25,Oosterloo+25}.

\subsection{VLA targets for this work}
\label{our_sample}

Out of the 39 galaxies in the SUPER Survey, we selected three (CID-346, XID-36, and J1333+1649) for VLA (Very Large Array) radio observations. These targets were selected as they represent the only galaxies from the SUPER Survey that are also part of the James Webb Space Telescope/Mid-Infrared Instrument Cycle 1, Program 2177 (PI: Mainieri) and thus were prioritized for new dedicated VLA observations. JWST enables for the first time to trace warm and hot molecular gas at Cosmic Noon using rest-frame near-infrared H$_{2}$ emission lines \citep{Kakkad+25}. In Appendix \ref{molecular_gas}, we briefly describe the molecular gas data for CID-346 \citep[already presented in][]{Kakkad+25} and place it in the context of our results in Section \ref{sec:Spatial_alignment}. The selection strategy aimed to investigate the connection between radio emission and gas outflows across different phases by complementing JWST and SINFONI data with radio observations.

In this paper, we focus on the connection between ionized gas and radio emission. These objects exhibit high W$_{80}$ values \citep[ranging from 1000 to 2500\kms\ for the \oiii\ emission line;][]{Kakkad+20, Tozzi+24}, which are associated with AGN-driven outflows, making them ideal prototypes for studying the interplay between radio emission and ionized gas outflows at cosmic noon. The three JWST targets were selected from the SUPER sample to span a broad range of bolometric luminosities, L$_{bol} \sim 10^{45.7}-10^{47.9}$ erg s$^{-1}$, as shown in Figure \ref{fig:sample}, as well as a wide range of mass outflow rates derived from \oiii\ emission, ranging from 0.6 to 1200 M$_{\odot}$ yr$^{-1}$ \citep{Kakkad+20,Tozzi+24}. They are also representative of a broad range of q$_{24 obs}$ and radio power at 1.4 GHz of the parent sample. The properties of these quasars are presented in Table \ref{table:sample}. 

The q$_{24obs}$ parameter is useful for identifying excess in radio emission, above that expected from purely star formation. The limitation of an analysis on q$_{24obs}$ is that it does not account for the AGN contribution to the 24 $\mu\text{m}$ luminosity. As a result, a source could have an artificially enhanced 24 $\mu\text{m}$ emission, potentially placing it within the star-forming region even if its emission is influenced by AGN activity. To take into account the increase in q$_{24obs}$ that could be produced by the AGN emission and address this issue, \citet{Circosta+18} estimated the average AGN contribution to the total flux at 24 $\mu\text{m}$, then this value was subtracted from q$_{24obs}$. The central panel of Figure \ref{fig:sample} shows the values of q$_{24obs}$ after the \citet{Circosta+18} q$_{24obs}$ correction versus redshift for the entire SUPER Survey sample. The solid line represents the typical values for a star-forming galaxy \citep[M32; from][]{Bonzini+13}, while the dashed lines indicate the $\pm 2\sigma$ dispersion, which marks the locus of star-forming galaxies \citep[see][]{Circosta+18}. Galaxies below this region exhibit excess radio emission; thus, our three sources are classified as radio-normal (CID-346), moderately radio-excess (XID-36), and strongly radio-excess (J1333+1649), as shown in Figure \ref{fig:sample}.  In addition to q$_{24obs}$, for CID-346 and XID-36, we have the FIR luminosities from \citet{Circosta+18}, which are plotted against the radio power in the right panel of Figure \ref{fig:sample}. The solid line in this plot represents the expected relation between FIR luminosity and radio power for star-forming galaxies. This relation is based on the works of \citet{Magnelli+15} and \citet{Bell+03}.  The positions of CID-346 and, in particular, XID-36 are below the expected relation between FIR luminosity and radio power. This deviation indicates an excess of radio emission relative to typical star formation, likely due to the presence of an AGN, whose jets or processes associated with the central nucleus may be contributing to the observed radio emission.  The combined analysis of the q$_{24obs}$ and radio-FIR correlation suggests that our three VLA targets show an excess of radio emission related to the presence of an AGN.

\section{Radio and ionized gas data}
\label{sec:radio}

The VLA imaging observations of the three targets were conducted under proposal ID 23A-074 (PI: S. Dougherty) using the A and B-arrays in C band at 6.2 GHz, providing a resolution of approximately 0.3\arcsec\ (2.5 kpc) and 1\arcsec\ (8.3 kpc), respectively (see beam size in Table \ref{table:observations}). The highest resolution (A-array) was chosen to match the spatial resolution of the spatially resolved spectroscopy from SINFONI, and thus a direct comparison between radio structures and ionized gas features can be made. The lower resolution (B-array) data at the same frequency were chosen to search for large-scale structures  ($\gtrsim$10\,kpc). The range of resolutions of the VLA images also approximately matches those of the mid-infrared spectroscopy from JWST, which will allow a future resolution-matched comparison. The description of the VLA observations and data reduction can be found in Section \ref{sec:vla_observations} and \ref{sec:vla_data_rededuction}.  In addition to the VLA observations in the C band, we incorporated some relevant radio data from the literature, as detailed in Section \ref{sec:radio_literature}. The maps for ionized gas, such as flux, centroid velocity, and \weighty, used in this study for the three quasars were obtained from \citet{Kakkad+20} for J1333+1649 and CID-346, and from \citet{Tozzi+24} for XID-36. We provide a brief description of how these maps were generated in Section \ref{sec:ionized_gas}; for a detailed description, we refer to the studies cited above. 

\subsection{VLA observations}
\label{sec:vla_observations}
The C band VLA observations in both A-array and B-array configurations, for all three targets, were undertaken in 2023 (exact dates in Table~\ref{tab:obs}). For the two  fainter targets (which required the longer on-source times) the observations were split across two sessions, to assist with scheduling. For the brightest target, J1333+1649, only one session was required in each array configuration. The observations blocks included initial exposures on flux/bandpass calibrators (3C286, 3C138, and 3C48), followed by regular slewing between nearby phase calibrator exposures and target exposures. The phase calibrators used, and the total on-target integration times are presented in Table \ref{tab:obs}.

\subsection{VLA data reduction and imaging}
\label{sec:vla_data_rededuction}
The data were reduced using VLA-supplied calibration scripts. We used the standard VLA {\sc casa} Calibration Pipeline with {\sc casa} version 6.4.1 \citep{casa}. Following the full calibration steps, we performed manual flagging, within the {\sc casa} tool ``plotms,'' to remove any frequency channels with residual bad data. Imaging was performed on the targets (after separating them from the full measurement sets) using the {\sc casa} task ``tclean,'' with the Multi-Frequency Synthesis mode. For all A-array images, a 125$\times$125 arcsec region was imaged with a pixel scale of 50 mas. In the B-array configuration, the imaged regions varied from 260$\times$260 arcsec to 425$\times$425 arcsec depending on the target, with the corresponding pixel scales detailed in Table \ref{table:observations} for each source. 

For the imaging process, we applied the Briggs weighting scheme to the baselines, exploring a range of robustness parameter values (between -1 and +1) for each target (see Table \ref{table:observations}), to explore the balance between resolution and sensitivity per beam. We chose a Briggs robust weighting parameter of 0 as a good compromise between sensitivity and resolution for most images, as shown in Table \ref{table:observations}. However, for the A-array images of CID-346 and XID-36, we adopted a Briggs weighting parameter of +1. This choice was motivated by the need to enhance the sensitivity of the images, thereby improving the detection of possible extended, low-surface-brightness structures in these sources. In our case, this trade-off allowed us to better visualize faint extended structure while still preserving sufficient angular resolution for our scientific analysis. For the very bright source, J1333+1649, additional self-calibration steps were performed during the imaging process. This involved two rounds of phase-only self-calibration, followed by one round of phase-plus-amplitude self-calibration. Table \ref{table:observations} includes the details of the final images for the three targets, which  can be found in Appendix \ref{Appendix_radio_images} for both the A and B arrays. 

\begin{figure*}
	\includegraphics[width=\textwidth]{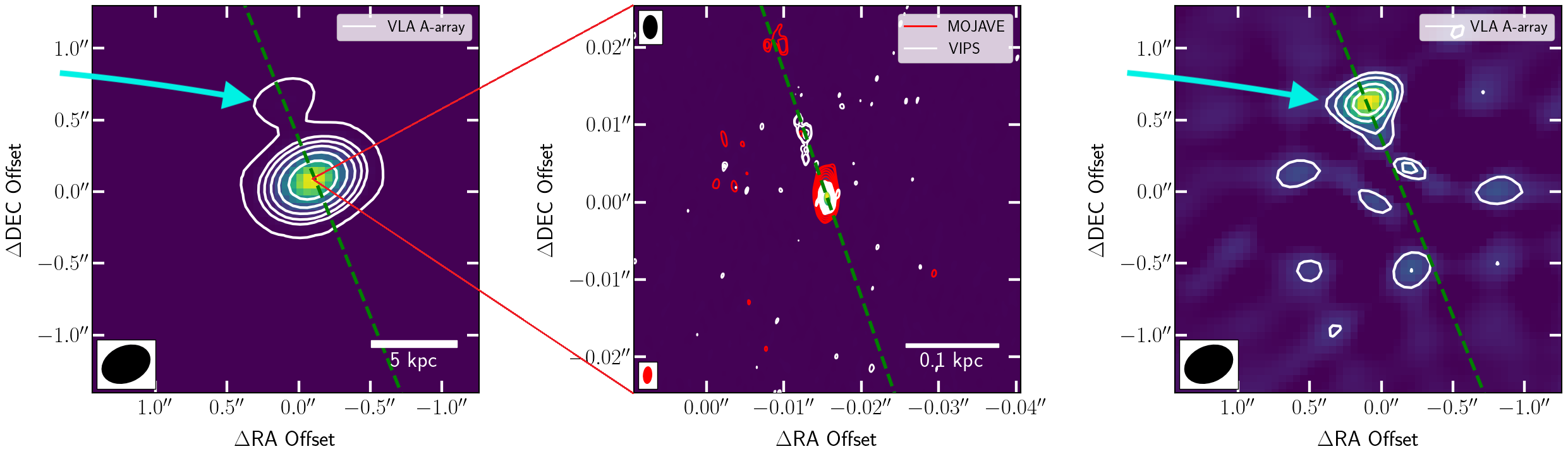}
    \caption{Radio maps for J1333+1649. The beam is represented by a ellipse on each image. Left: VLA A-array 6.2 GHz radio map with contours plotted at levels of [36, 512, 1024, 2048, 4096, and 8192]$\sigma$.  The dashed line indicates the position angle of the radio extended structure. Middle: MOJAVE 8.1 GHz  radio map (background map) with MOJAVE 8.1 GHz and VIPS 5.0 GHz radio contours at [3, 4, 8, 16, 32, 64, 128, and 256]$\sigma$.  The dashed line indicates the PA of the VIPS radio extended structure, which is similar to that of MOJAVE. Right: VLA A-array residual map from {\sc casa} after subtracting the central source with contours plotted at [16, 32, 64, and 96]$\sigma$. The cyan arrow indicates the extended radio structure.}
    \label{fig:J1333_radio}
\end{figure*}

\begin{figure*}
	\includegraphics[width=\textwidth]{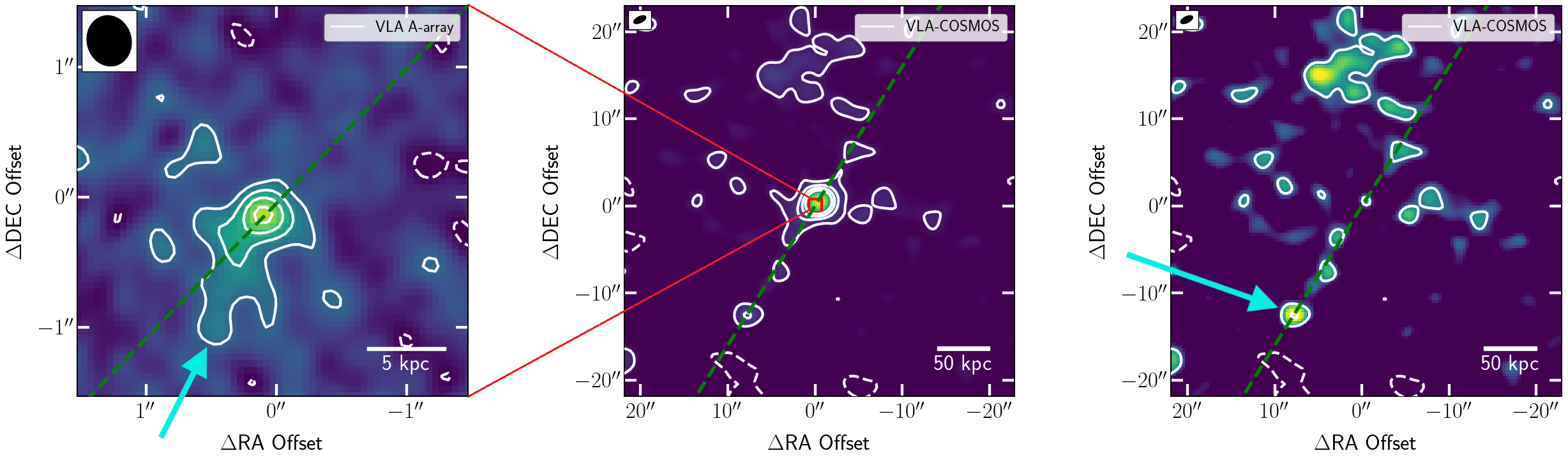}
    \caption{ Radio map contours and residual map of images of CID-346. The beam is represented by a ellipse on each image. The dashed lines indicate the position angles of the extended structure. Left: VLA A-array with contours [-2, 2, 4, 6, and 8]$\sigma$. Middle: VLA-COSMOS radio map with contours plotted at [-2, 2, 4, 6, and 8]$\sigma$. Right: VLA-COSMOS residual map after removing central source with {\sc casa} with contours plotted at [-2, 2, and 4]$\sigma$.  The VLA A-array observations were conducted at 6.2 GHz, while the VLA-COSMOS survey was performed at 1.4 GHz. The cyan arrows indicate the extended radio structure.}
    
    \label{fig:cid346_radio}
\end{figure*}

\begin{figure}
 	\includegraphics[width=0.40\textwidth]{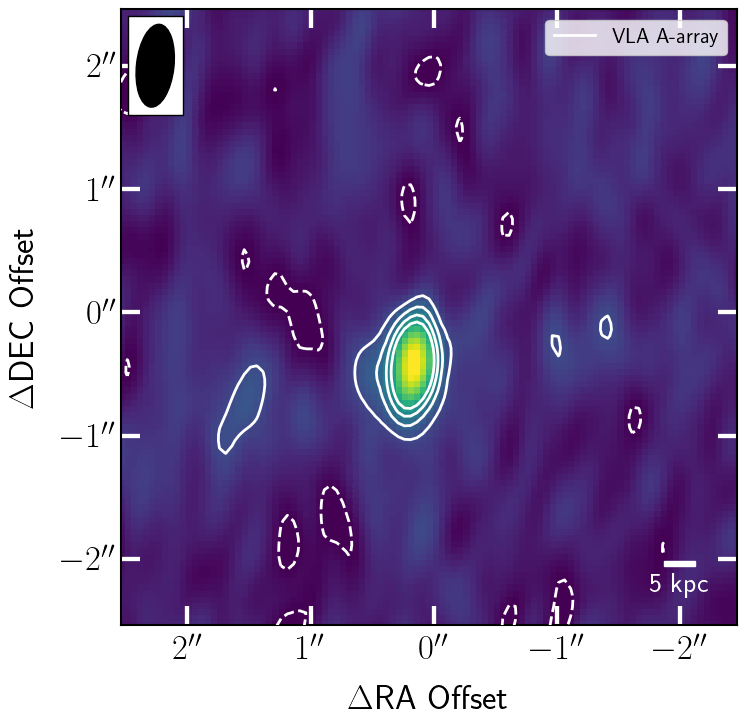} 
      \caption{Radio map for XID-36. The contours at [-2, 2, 4, 6, and 8]$\sigma$ show the VLA A-array radio data. The beam for the radio data is represented by an ellipse. The VLA A-array observations were conducted at 6.2 GHz. There is no strong evidence for extended radio structures in this source.} 
    
    \label{fig:XID36_radio}
\end{figure}

\subsection{Radio data from literature}
\label{sec:radio_literature}

 For J1333+1649, these images allowed us to investigate radio structures at significantly smaller scales than the $\sim$ 1 kpc achieved by our A-array VLA observations. Therefore, we also utilized available data from the VLBA Imaging and Polarimetry Survey \citep[VIPS;][]{Helmboldt+07} and the Monitoring of Jets in Active Galactic Nuclei with VLBA Experiments \citep[MOJAVE;][]{Lister+18}.  VIPS is a radio survey conducted at 5 GHz and 15 GHz using the Very Long Baseline Array (VLBA), covering approximately 1000 potential AGN. For J1333+1649, we used the 5 GHz radio observation, which has a beam size of approximately 0.20 mas $\times$ 0.3 mas with a position angle of 0$^\circ$. MOJAVE is a project that focuses on studying parsec-scale jets in AGN, which also employs VLBA imaging \citep{Lister+18}. We used the 8.1 GHz radio data from MOJAVE, which provides a beam size of 1.1 mas $\times$ 2.1 mas with a position angle of -4.2$^\circ$. Since CID-346 exhibits an extended structure in the VLA A-array image (Section \ref{cid346_radio_data}), we explored additional data for this object at different scales to determine whether extended structures are present at other resolutions.
Thus, for CID-346, we analyzed the 1.4 GHz VLA data from the COSMOS survey \citep[Cosmic Evolution Survey;][]{Schinnerer+04}, which has a beam size of 1.5\arcsec$\times$0.7\arcsec (12.4$\times$5.8 kpc) with a position angle of -63.5$^\circ$.  It is worth noting that no archival data were found that could help in interpreting the results for XID-36.

\subsection{Ionized gas}
\label{sec:ionized_gas}
In this study, we focus on comparing the spatial distribution and kinematic structure of the \oiii\ gas with the radio images. To accomplish this, we utilize previous kinematic analyses of our targets conducted by \citet{Kakkad+20} and \citet{Tozzi+24}. The \oiii\ emission line was fitted by \citet{Kakkad+20} and \citet{Tozzi+24} using Gaussian functions.  In addition, non-parametric measurements were adopted for the \oiii\ emission line, from which properties such as \vten\ and \weighty\ were obtained. \weighty\ represents the velocity width containing 80\% of the total emission line flux and is commonly used to identify outflows \citep[see][]{Liu+13,Wylezalek+20,Kakkad+20,Gatto+24}. \vten\ corresponds to the velocity at the tenth percentile of the overall emission line profile \citep{Kakkad+20}.  \citet{Kakkad+20} also demonstrated  that the ionized gas in all the type-1 AGN in the SUPER Survey with detected \oiii\ show the presence of ionized outflows, considering \weighty>600 \kms\ as a tracer of outflows. According to \citet{Kakkad+20}, the \weighty>600 \kms\ threshold was chosen after comparing the SUPER sample with mass-matched galaxy samples at both low and high redshifts from other surveys, such as MaNGA \citep[Mapping Nearby Galaxies at Apache Point Observatory;][]{Bundy+2015} and KLEVER \citep[K-band Multi Object Spectrograph Lensed Emission Lines and VElocity Review;][]{Curti+2020}. Therefore, in this study, we assume that \weighty\ values exceeding 600 \kms\ for our  quasars (both type-1 and type-2) are associated with AGN-driven outflows.
Although this method identifies the most extreme outflow regions, we do not rule out that there are less extreme outflows in regions with \weighty\ below this threshold, which may still be energetically important \citep{Ward+24}.

\subsection{Alignment of radio and SINFONI data}
\label{sec:alignment}
Since we aim to examine the potential connections between the extended radio structure and the ionized gas, it is crucial to ensure that the astrometry of the SINFONI data has been properly corrected. To achieve this, we initially utilized the coordinates RA and Dec for J1333+1649 and for CID-346, obtained from Gaia Archive \citep{GaiaCollaboration+16, GaiaCollaboration+23}. These coordinates were then matched with the central pixel values determined by fitting a 2D Gaussian to the continuum maps (shown in Figure \ref{cont_map}) derived from the SINFONI H-band observations. However, following the approach of \citet{Kakkad+25}, we also considered re-aligning the data by matching the peak of the radio emission to the peak of the H-band continuum for J1333+1649 and CID-346, rather than relying on the Gaia coordinates. This alternative alignment was motivated by three considerations: first, it ensures consistency with the method adopted in the published work of \citet{Kakkad+25}; second, it addresses the small misalignment between the continuum and radio emission when Gaia astrometry is applied; and third, although the required shift is minor (of the order of $\sim$1 pixel), it can significantly improve the overall alignment, particularly for J1333+1649. For XID-36, the H-band continuum is not detected by \citet{Tozzi+24}, as the spectra, being those of a distant type-2 AGN, do not exhibit a strong continuum from either the AGN or stellar components. This target is also not part of the GAIA catalog; thus, the astrometric correction (or radio–continuum alignment) was not performed for this object. In Section \ref{xid36_gas}, we discuss the results for XID-36 and note that the absence of astrometric correction does not impact our conclusions, as no resolved structures beyond the central emission are identified in the radio data.

\section{Radio analysis}
\label{sec:radio_results}

We followed different approaches to analyze each of the observed AGN, since we also have varied radio data from the literature and they cover different radio flux regimes. To investigate the connection between radio emission and ionized gas, we first examined whether the radio images displayed any evidence of extended structures. Initially, each image was inspected visually using {\sc SAOImageDS9} \footnote{\url{https://sites.google.com/cfa.harvard.edu/saoimageds9}} \citep{ds9}.
Subsequently, we generated radio maps overlaid with contour levels scaled relative to the noise. These contours help establish the presence or not of structures based on their significance level compared to the noise, enabling the detection of extended features that might not be readily visible. All our VLA radio images are displayed in Figure \ref{fig:large_vla} (see Section \ref{sec:vla_observations}), but we focus on the most interesting features with the maps presented in Figures \ref{fig:J1333_radio}-\ref{fig:XID36_radio}, including relevant archival data (Section \ref{sec:radio_literature}).  

\subsection{J1333+1649}
For the source with the highest radio emission excess, i.e., J1333+1649, the radio contours for VLA A-array, MOJAVE and VIPS are shown in Figure \ref{fig:J1333_radio}.  Residual maps are crucial for highlighting morphological structures that are not well described by a two-dimensional  Gaussian fit. These maps can reveal potential extended features, such as possible jets. For this target, we removed the extremely bright point source, to look for extended structures. Therefore, we created a residual map after subtracting the point source. We did this, for the VLA A-array, by performing a two-dimensional Gaussian fit from the peak of the radio source emission using the {\sc casa}'s (version {\sc casalith} 6.6.4.34) {\sc imfit} function. The residual map for VLA A-array is also presented in Fig. \ref{fig:J1333_radio}. The VLA A-array reveals an extended structure ($\sim$ 0.5\arcsec or $\sim$ 4.2 kpc) to the northeast of the emission peak with 36$\sigma$ above the noise level. This structure becomes even more apparent in the VLA A-array residual map, where a blob with higher intensity is observed to the northeast with more than 16$\sigma$ significance from the noise on a scale of $\sim$2 kpc, the residual suggests a collimated and unidirectional structure. 
The VLA B-array provides the lowest-resolution image for J1333+1649 (Figure \ref{fig:large_vla}). In this case, the source is described by the {\sc casa} fitting process as an unresolved source. There is no evidence of an extended structure detectable within the resolution limits of this B-array image and no evidence of large-scale structure on $\gtrsim2$ kpc scales.

The existence of this structure in the VLA A-array motivated us to examine the available VLBI (Very-long-baseline interferometry) radio images, which offer much higher resolution. The higher-resolution images from VIPS and MOJAVE (central panel in Figure \ref{fig:J1333_radio}) reveal a more detailed view of the emission on smaller scales, with an elongated  structure observed up to 0.01\arcsec\ (0.08 kpc) from the nucleus for VIPS and 0.02\arcsec\ (0.17 kpc) for MOJAVE. As shown by the dashed line in central panel of Figure \ref{fig:J1333_radio}, these features appear to extend from the inner jet-associated knots, further supporting their connection to the larger-scale structure. The detection of this extension is confirmed with at least 3$\sigma$ significance values for MOJAVE and VIPS. According to \citet{Panessa+19}, radio structures resolved on mas scales, such as those we observed in our J1333+1649 with MOJAVE/VIPS, indicate more compact sources and are associated with jets. The orientation of these structures across all three images (VLA, MOJAVE and VIPS) was quantified by obtaining the position angle, which was measured anti-clockwise relative to the north direction of the images. The PA values were determined by defining an axis represented by a line connecting the spaxel corresponding to the peak of the central emission and the spaxel associated with the peak of the extended off-nuclear emission, identified at the highest sigma level. 

For the MOJAVE dataset, the PA was determined considering the structure extending up to 0.02\arcsec, while for the VIPS dataset, the extended structure was considered up to 0.01\arcsec. The position angles of the jet-like structure are presented in the Table \ref{table:pa_combined}. The position angles for the VLA A-array and for VIPS are also represented by the dashed line in Figure \ref{fig:J1333_radio}. Given the relative large beam size in the VLA A-array image, the structure does not appear very well collimated. Therefore, there is an inherent uncertainty in the position angle, which can be estimated by considering the extended structure at the lowest $\sigma$ level. These uncertainties are also presented in the Table \ref{table:pa_combined}. By combining the three images our analysis reveals a jet extended along a consistent axis with a position angle of $\sim10^{\circ}$, from $\sim$80 pc to $\sim$5 kpc. This result marks the first time that the jet in this source has been identified as extending on host-galaxy scales, which is particularly relevant for understanding its interaction with the ISM. The implications of this result are further discussed in Section \ref{sec:J1333} and \ref{sec:Radio_outflow}. 

It is particularly noteworthy that in J1333+1649 we observe jet-like structures on different spatial scales and at multiple frequencies. The detection of extended radio emission on scales of $\sim$5 kpc with the VLA A-array, together with compact jet features revealed by MOJAVE/VIPS, may indicate the possibility of recurrent AGN activity. A possible discrepancy between the dynamical timescales of the large-scale emission and the parsec-scale jet suggests that they could originate from distinct activity episodes. This interpretation remains preliminary, but is consistent with the restarted-jet scenario discussed for other radio sources \citep[e.g.,][]{Brocksopp+07,Shulevski+15,Kukreti+23,Rao+23}, where large-scale emission is interpreted as a signature of past activity while compact jets are associated with more recent nuclear activity. It has been extensively investigated that AGN activity proceeds through different cycles, with active phases followed by periods of quiescence and subsequent reactivation. For radio AGN, these different phases can leave observable imprints: multiple radio features may be detected along the same axis of propagation, each corresponding to a distinct episode of nuclear activity \citep[see][for a review]{Morganti+17}.

\subsection{CID-346}
\label{cid346_radio_data}
For the quasar CID-346, the VLA B-array image (central panel of the Figure \ref{fig:large_vla}), corresponding to the lowest resolution of our VLA data, exhibits emission that is well described by a two-dimensional Gaussian fit, with no significant residual signal detected and with no clear signs of extended structures. In contrast, the higher-resolution radio image, obtained with the VLA A-array (Figure \ref{fig:cid346_radio}), reveals an extended structure to the southeast of the central emission, detected at a significance level of at least 4$\sigma$, with an even more extended  contiguous structure at the 2$\sigma$ level.  The position angle measured  for the structure detected at a significance level of 4$\sigma$ is 130$^{\circ} \pm 17 ^{\circ}$.  However, the {\sc casa} software cannot fit the source with a good two-dimensional Gaussian model, as the fit incorporates the extended structure as part of the 2D Gaussian description, since the extended structure is not so bright. As a result, the production of a residual map for this image is not feasible. Nevertheless, the fit provides a consistent position angle of 146 $^{\circ}$$\pm$29$^{\circ}$,  confirming that the source is extended in this direction.

For this object, a 1.4 GHz VLA image is also available, which has a lower resolution than our VLA A and B-array observations. Due to its lower frequency and resolution, this image may be more sensitive to diffuse structures and/or emission associated with lower-frequency components. This image (central panel of Figure \ref{fig:cid346_radio}) reveals knots extending up to $\sim$13\arcsec\ ($\sim$107 kpc) toward the southeast, detected with a significance of 2$\sigma$ and 4$\sigma$. Although observed predominantly with 2$\sigma$ significance, the extended radio structure  is approximately in the same direction as the extended structure from the VLA A-array.  After obtaining the residual map for COSMOS (right panel of Figure \ref{fig:cid346_radio}) we observe a slight excess emission in the central region with a significance of 2$\sigma$, which is connected to the emission observed toward the southeast at an angle of 148$^{\circ} \pm 6 ^{\circ}$ determined considering the most extended structure detected ($\sim$13\arcsec\ toward the southeast).  These position angles are summarized in Table \ref{table:pa_combined}. Furthermore, the 1.4 GHz VLA-COSMOS image reveals a 2$\sigma$ level structure to the north (extending up to 20\arcsec), which may be associated with the central source. In the case of CID-346, as discussed for J1333+1649, the coexistence of extended radio emission detected with VLA-COSMOS and compact structures observed on smaller scales with the VLA A-array likewise suggests a restarted-radio scenario, in which the large scale emission traces past activity while the core reflects more recent nuclear activity. Despite the detection of extended radio morphology in CID-346 with both the VLA A-array and VLA-COSMOS, we cannot confirm that this emission is associated with a jet, although we consider this a possible hypothesis.

\subsection{XID-36}

XID-36 is the only AGN in our study that does not exhibit extended radio structures  (see Figure \ref{fig:XID36_radio} ). Neither the A-array (Figure \ref{fig:XID36_radio} and Figure \ref{fig:large_vla}) nor the B-array (Figure \ref{fig:large_vla}) show evidence of extended radio structures on the scales probed by our observations (several kiloparsecs). In the VLA A-array data,  {\sc casa} was unable to deconvolve the source from the beam; the source is unresolved. Thus, we did not obtain the radio PA for the VLA  A-array image. For the VLA B-array, the source is resolved and shows a PA of 15$^{\circ} \pm 10^{\circ}$. This measurement should be interpreted with caution, as the B-array image may be tracing additional, more extended emission that is resolved out in the higher-resolution A-array image. Given the relatively large beam and the inferred compact size of the source, we do not further interpret this PA. In the remainder of this work, we focus only on sources with clearly defined extended radio structures, such as J1333+1649 and CID-346.

\begin{figure*}
	
	\includegraphics[width=\textwidth]{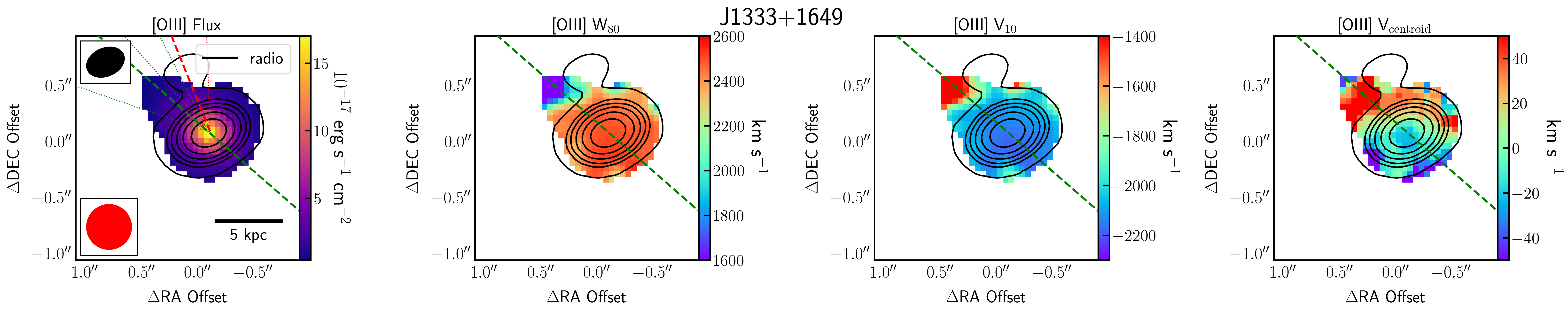}
    \\
    	\includegraphics[width=\textwidth]{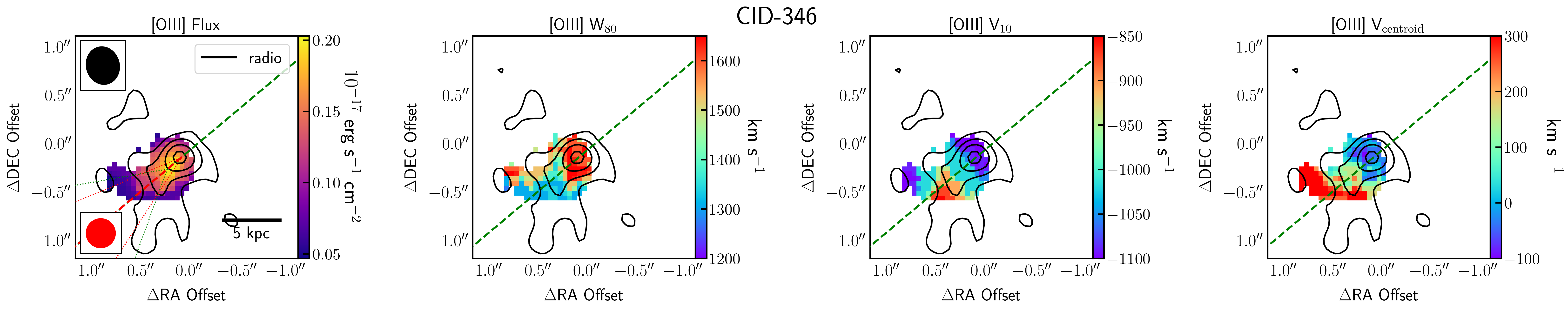}
 \\
 \includegraphics[width=\textwidth]{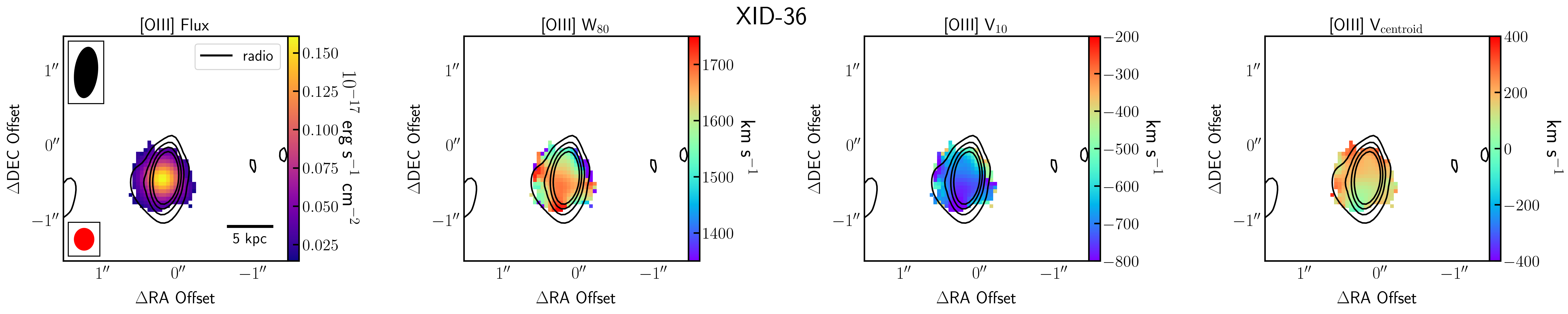}

    \caption{\oiii\  maps for our three targets. The PSF and beam for the ionized gas maps and radio image are represented by a red and a black ellipse, respectively. The green dashed line indicates the position angle of  \oiii\ emission, while the red dashed line indicates the radio PA. The dotted lines indicate the uncertainties of the PAs. Top: Maps for J1333+1649. The contours plotted at [36, 512, 1024, 2048, 4096, and 8192]$\sigma$ show the VLA-A array radio data. Middle: Maps for CID-346. The contours at [2, 4, 6, and 8]$\sigma$ show the VLA-A array radio data. Bottom: Maps for XID-36.  The  contours at [2, 4, 6, and 8]$\sigma$ show the VLA-A array radio data. } 
    
    \label{fig:J1333_oiii_radio}
\end{figure*}

\section{Spatial connection between radio and ionized gas emission}
\label{sec:radio_gas}

The maps obtained by \citet{Kakkad+20} and  \citet{Tozzi+24} for the ionized gas (as described in Section \ref{sec:ionized_gas}) are presented in Figure \ref{fig:J1333_oiii_radio} for our targets.  The maps are presented with a signal-to-noise (S/N) cut larger than two for CID-346 and  J1333+1649. For XID-36, they are shown with a signal-to-noise cut of S/N > 3.0. Since the \oiii\ measurements from SINFONI are matched in resolution to the VLA A-array radio data, we can directly compare the spatial distributions and properties derived from both datasets. Therefore, Figure \ref{fig:J1333_oiii_radio} also include contours for the radio emission.  In the following sections, we present our findings for each of the three targets individually (Sections \ref{sec:J1333}- \ref{xid36_gas}).

\subsection{J1333+1649}
\label{sec:J1333}
The \oiii\ emission of this target extends to regions beyond 6 kpc northeast of the center. The analysis of the spaxel-by-spaxel maps for the kinematics supports the presence of outflows in this target, with \vten\ indicating blueshifted velocities across the entire field, reaching speeds of up to -2100 \kms\ in the same regions where \weighty\ reaches its maximum values of $\sim$2600 \kms. An increase in \vten\ values, reaching -1400 \kms, occurs at the location where we observe the extended emission to the northeast of the center. Furthermore, in this region, we also observe a decrease in \weighty, reaching $\sim$1600 \kms. We determined the morphological position angle for the ionized gas considering the direction of extended structure to the northeast of the center (see the green dashed line in Figure \ref{fig:J1333_oiii_radio}). The PA is measured relative to the north direction of the image, anticlockwise. The PA value is $49^{\circ} \pm22$, which is also presented in Table \ref{table:pa_combined}.  In order to check the connection between the radio emission and the ionized gas, we overlay the VLA A-array radio contours for J1333+1649 in Figure \ref{fig:J1333_oiii_radio}. Both the radio blob and the extended \oiii\ emission are observed in a northeasterly direction, extending to roughly the same radial distance ($\sim$2 kpc). It is important to note that the spatial resolution of the SINFONI and VLA A-array data are well matched, minimizing the impact of beam smearing on the comparison.  Although they show a projected misalignment of approximately 27$^{\circ}$ between their position angles, when taking into account the uncertainties listed in Table \ref{table:pa_combined}, the orientations are consistent inside these uncertainties. We also determined the kinematical PA (see Table \ref{table:pa_combined}) for the ionized gas to compare its orientation with that derived from the ionized gas morphology. The kinematical PA was measured from velocity centroid map using the PaFit package \citep{Krajnovic+06}. For J1333+1649, the uncertainty in the kinematic PA is larger, but the PA value is closer to the radio PAs. In this case, the kinematic and morphological PAs are marginally consistent within the uncertainties. Overall, these results indicate that the morphological and kinematical position angles are consistent and show no significant discrepancies within the measurement errors. Therefore, the extended radio structure appears to be aligned with the location of the sudden drop in \weighty\ and a sudden increase in \vten.

\subsection{CID-346}
The \oiii\ flux map displays emission in regions extending beyond 6 kpc, as shown in Figure \ref{fig:J1333_oiii_radio}. Similarly to the previous source, this quasar exhibits blueshift velocities across the entire field of view (see \vten\ in Figure \ref{fig:J1333_oiii_radio}), although with a smaller amplitude than the previous target, reaching approximately -1100 \kms. The \weighty\ values also suggest that the kinematics are primarily dominated by outflows, with velocities consistently exceeding 1300 \kms\ and reaching a peak of 1650 \kms. In the maps shown in Figure \ref{fig:J1333_oiii_radio}, we also overlay the contours of the VLA A-array radio emission. The extended radio structure is closely aligned with the direction of the \oiii\ flux, as demonstrated by the position angle listed in Table \ref{table:pa_combined}. In particular, considering the 4$\sigma$ radio contour, we measure a position angle of 130$^{\circ} \pm 17^{\circ}$, which is in excellent agreement with the ionized gas morphological orientation of 130$^{\circ} \pm 30^{\circ}$. The kinematic PA is 158$^{\circ} \pm 9^{\circ}$, and the uncertainty in the kinematic PA is smaller than that of the morphological PA, with both measurements being broadly consistent within the uncertainties. We also note the presence of a possible extended structure in the VLA-COSMOS map. However, there is evidence that the radio structure changes PA from $\sim$130$^{\circ}$ to $\sim$150$^{\circ}$ (at the location of the extended ionized gas) from both VLA 6.2 GHz image (considering the  2$\sigma$ contour) and the VLA-COSMOS 1.4 GHz image. It is important to note that the spatial resolution of the SINFONI and VLA A-array data are well matched for this target as well, ensuring that the comparison is not significantly affected by beam smearing. For CID-346, we also observe a drop in \weighty\ values and an increase in \vten\ in the region where the radio emission appears to change direction.

\subsection{XID-36}
\label{xid36_gas}
Figure \ref{fig:J1333_oiii_radio} also presents the flux and kinematic maps of the ionized gas in XID-36, overlaid with the radio emission contours from the VLA A-array.  We did not compare the position angles of the radio emission with those of the ionized gas, since the source is not resolved in the VLA A-array data. Both the \oiii\ and radio emission  appear to exhibit similar extents, but this apparent similarity remains inconclusive given the limitations imposed by the available angular resolution. The velocity maps indicate the presence of ionized gas outflows, with \weighty\ > 1350 \kms\ and \vten\ showing blue-shifted velocities ranging from -200 to -800 \kms. Previously, we noted that the absence of astrometric correction does not affect our conclusions for XID-36. This is because the radio data show no evidence of extended structures beyond the central/nuclear region. Therefore, any small astrometric uncertainties have no impact on the interpretation of its morphology or orientation. It is interesting to note that XID-36 is the only source in our sample, which does not show extended radio structures, and is also the only source which does not show extended ionized gas structures.

\begin{figure}
\centering

	\includegraphics[width=0.45\textwidth]{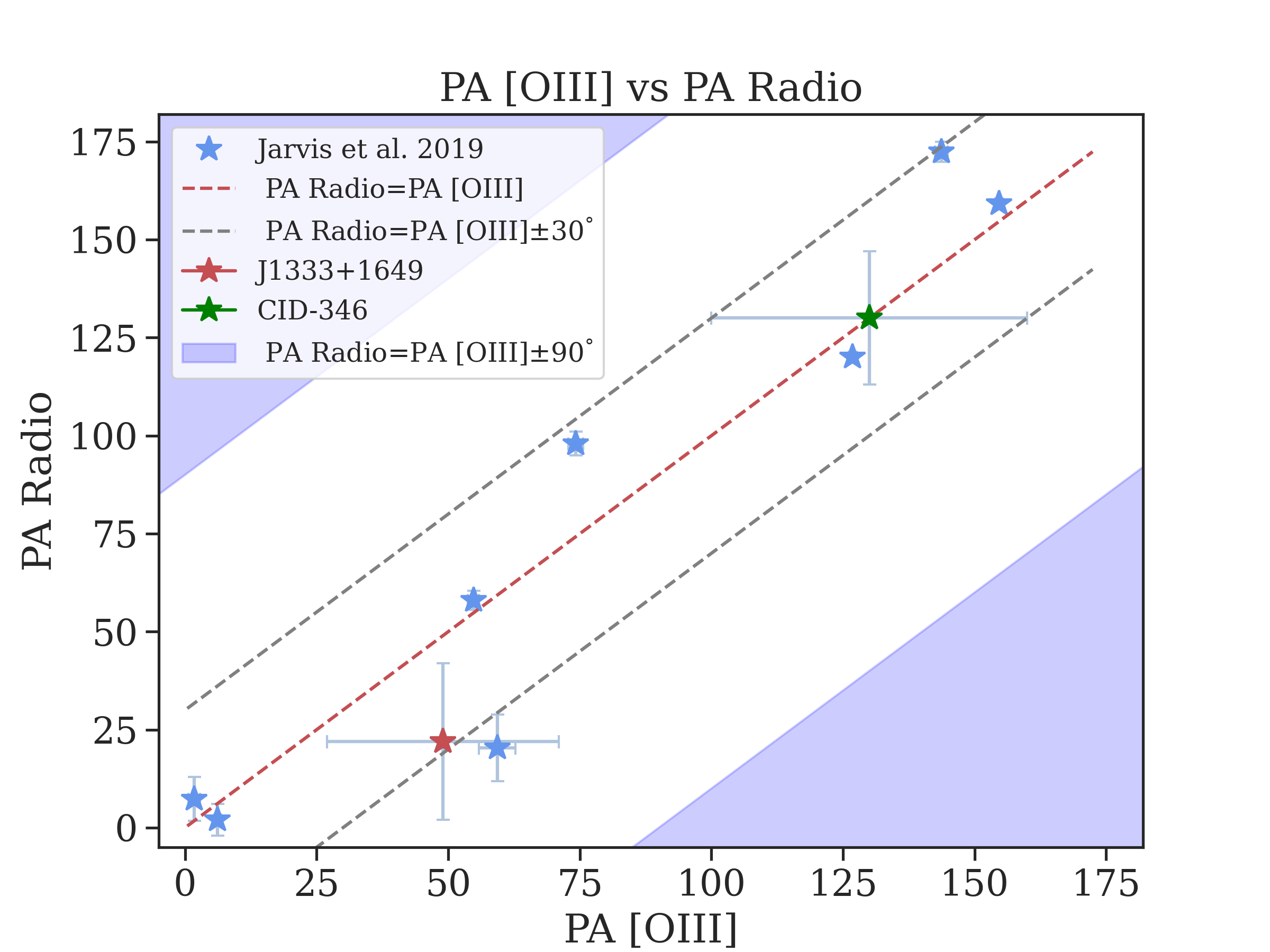}

    \caption{Position angle of the major axis of the \oiii\ and the radio emission for the galaxies in the \citet{Jarvis+19} sample and our targets. Sources located within the gray dashed lines are considered aligned (see Section \ref{sec:Spatial_alignment}). XID-36 does not show any extended structures, and the radio emission is poorly resolved; therefore, we do not have a reliable radio PA for this object. }
    \label{fig:pa_jarvis}
\end{figure}

\section{Connecting outflows with radio emission}
\label{sec:Radio_outflow}
In this section, we explore the connection between radio emission and ionized outflows in typical radiative AGN at cosmic noon and in more powerful radio sources, extending findings established at lower redshifts. We begin with a purely empirical analysis, comparing the spatial distribution of the radio emission and the ionized gas (Section \ref{sec:Spatial_alignment}), and then consider whether the energetics of the radio jets could plausibly account for the observed outflows (Section \ref{Energetic_Perspective}).

\subsection{ Spatial alignment with the ionized gas outflows}
\label{sec:Spatial_alignment}

A strong indication of a connection between \oiii\ and radio emission can be seen in Figure  \ref{fig:J1333_oiii_radio}, where the ionized gas emission is co-spatial with the radio structure, displaying comparable spatial extents. The only exception is CID-346, where radio emission appears significantly larger than ionized gas emission when considering the 2$\sigma$ contours. However, at the 4$\sigma$ level, the radio emission size closely matches that of \oiii.  As previously discussed, the kinematics of these galaxies are dominated by ionized gas outflows, primarily characterized by high \weighty\ values across the \oiii\ entire field of view. Consequently, the detected outflows are co-spatial with the region of observed radio emission. Interestingly, the most extreme ionized gas kinematics, as indicated by the highest \weighty\ values, tend to coincide with the peaks of the radio continuum emission in the VLA A-array images. In the case of J1333+1649,  higher-resolution radio observations from MOJAVE and VIPS reveal a compact jet on sub-kiloparsec scales ($\lesssim 1$ kpc), suggesting that it may be driving the ionized outflows detected in the system. This compact jet could be responsible for the elevated \weighty\ values, once it is aligned with the regions showing the most extreme kinematic signatures (see Figure  \ref{fig:J1333_oiii_radio}). These results are consistent with predictions from jet-ISM interaction simulations \citep[e.g.,][]{Mukherjee+16,Mukherjee+18, Cielo+18,Meenakshi+22}, which show that jets couple efficiently with the ISM, excavating cavities and driving outflows. 

Furthermore, there is a close alignment between the spatial distribution of the radio extended structures and the ionized gas emission in both targets that exhibit extended structures. Although the radio and gas PAs are not exactly identical in J1333+1649, the offset remain within a range that supports a connection between the radio and the ionized gas (Section~\ref{sec:J1333}). The position angles of the extended radio structures and the ionized gas are shown in Figure \ref{fig:pa_jarvis}, alongside with the comparison sample from \citet{Jarvis+19}, who identified a relationship between the spatial distribution of ionized gas and radio structures for quasars with $z<0.2$, as quantified by the position angle. By combining both samples (Figure \ref{fig:pa_jarvis}), we find that our galaxies follow the same trend observed by \citet{Jarvis+19}, with a strong correlation between the two position angles and with almost all galaxies displaying a PA offset of less than $30^{\circ}$. Furthermore, the Pearson test returns a correlation coefficient of 0.95  and a p-value of 10$^{-6}$ when combining both samples, indeed indicating a strong correlation between the radio PA and the ionized gas PA. In contrast with results (including ours) that suggest an alignment or only a small offset between the radio and ionized gas position angles, the MURALES \citep[MUse RAdio Loud Emission line Snapshot;][]{Balmaverde+22} survey at low redshift ( $z < 0.3$) shows that extended emission line regions are often found at large angles relative to the radio axis. The kinematical axis of the ionized gas on the kiloparsec scales typically forms a median angle of $\sim65^{\circ}$ with the radio axis, indicating frequent misalignments.

A connection between ionized gas and radio emission at cosmic noon was also found by \citet{Nesvadba+17} for a sample selected to represent the most powerful radio galaxies, with radio power in the range $10^{26.3}-10^{29.3}$ W Hz$^{-1}$ . Therefore, they are not representative of ``typical'' radiative AGN at this epoch (i.e., they are high redshift radio galaxies). The \citet{Nesvadba+17} sample shows a distribution of PA offsets between ionized gas and radio predominantly skewed toward values lower than 30$^{\circ}$. Their sample also exhibits highly disturbed kinematics, largely characterized as being due to outflows for a significant portion of the sample.  For our sample,  the position angles of ionized gas and radio data have offsets not exceeding 30$^{\circ}$.  

\citet{Nesvadba+17} suggests that these offsets $<$30$^{\circ}$ found in their samples may result from jet jittering (small and rapid changes in the jet direction) or possible jet precession, as well as from low-surface-brightness regions that affect angle determination and the broad lateral extent of emission-line regions perpendicular to the jet direction. The ionized gas emission exhibits a broadened morphology along its major axis, rather than a collimated structure, resulting in a significant spatial extent perpendicular to the axis. As a result, the morphology of the ionized gas is not dominated by a single, well-defined axis, which makes it challenging to determine a precise position angle for the gas distribution. This difficulty is further compounded by the presence of low-surface-brightness regions, which often extend irregularly and may not follow the same orientation as the brighter structures. These diffuse features introduce additional uncertainty when attempting to define the overall orientation of the emission. For J1333+1649, jet precession or jittering appears unlikely, as the radio position angles measured at both parsec scales (from VIPS and MOJAVE data) and kiloparsec scales (from VLA A-array observations) are remarkably consistent (see Table \ref{table:pa_combined}). Given that this target exhibits such consistent radio position angles, we also find no evidence of jet deflection. 

The source CID-346 exhibits particularly intriguing behavior. At the 4$\sigma$ confidence level, we find an excellent alignment between radio and ionized gas emission. However, when the 2$\sigma$ level is considered, a slight deviation in the radio orientation seems to emerge, which agrees with the direction seen in the VLA-COSMOS 1.4 GHz large-scale structure. This variation in the radio structure orientation could be the result of jet deflection due to the interaction with a dense cloud in the ISM,  jet precession, or to the low signal-to-noise. \citet{Jarvis+19}, for instance, found indications of interactions between radio jet structures and the warm ionized gas and reported possible deflections of the radio jet in some of their targets as a result of encountering gas clouds. Indeed, this scenario may be relevant for CID-346, which presents a particularly complex environment. Recent observations by \citet{Kakkad+25}, combining MIRI/MRS spectroscopy and NIRCam imaging, revealed the presence of hot molecular gas toward the southeast of the AGN position (see Appendix \ref{molecular_gas} and Figure \ref{h2_map}). The extended hot molecular gas emission appears to be in the direction (southeast) of the extended ionized gas detected by \citet{Kakkad+20}. Additionally,  two satellite galaxies have been identified in the southeastern vicinity of CID-346. Furthermore, in this direction, extended CO(3-2) emission is detected \citep{Kakkad+25}, although it is not co-spatial with the radio emission. 

The interaction between the jets and the surrounding environment has been proposed as a mechanism responsible for producing jet deflections and bent morphologies \citep[e.g.,][]{Couto+13,Rawes+18,Maccagni+21,Park+24}. In this context, \citet{Holden+24} observed a bend in the small-scale radio structure, which is attributed to the interaction of the jet with dense molecular gas, leading to its deflection. Moreover, \citet{Wang+25} reports that companions may serve as the triggering mechanism for a powerful radio-loud AGN at $z \sim 3.5$ and further suggests that companion systems could play a role in deflecting the jet. In this scenario, given the availability of both ionized and molecular gas, the jet may not only be deflected by dense gas structures, but could also deplete the CO-based molecular gas reservoir, while simultaneously enhancing the emission from hot molecular gas, as suggested by \citet{Kakkad+25}. However, we cannot rule out an alternative interpretation in which the observed outflows result from accretion disk winds or radiative driving, with the resulting shocks producing the observed radio emission \citep[e.g.,][]{Zakamska+14,Nims+15}. Due to the limited spatial resolution of the available observations for CID-346 (compared to the archival data for J1333+164),  we are unable to identify collimated structures that could be unambiguously associated with a radio jet.

\subsection{An energetic perspective}
\label{Energetic_Perspective}
\begin{figure}
	\includegraphics[width=0.45\textwidth]{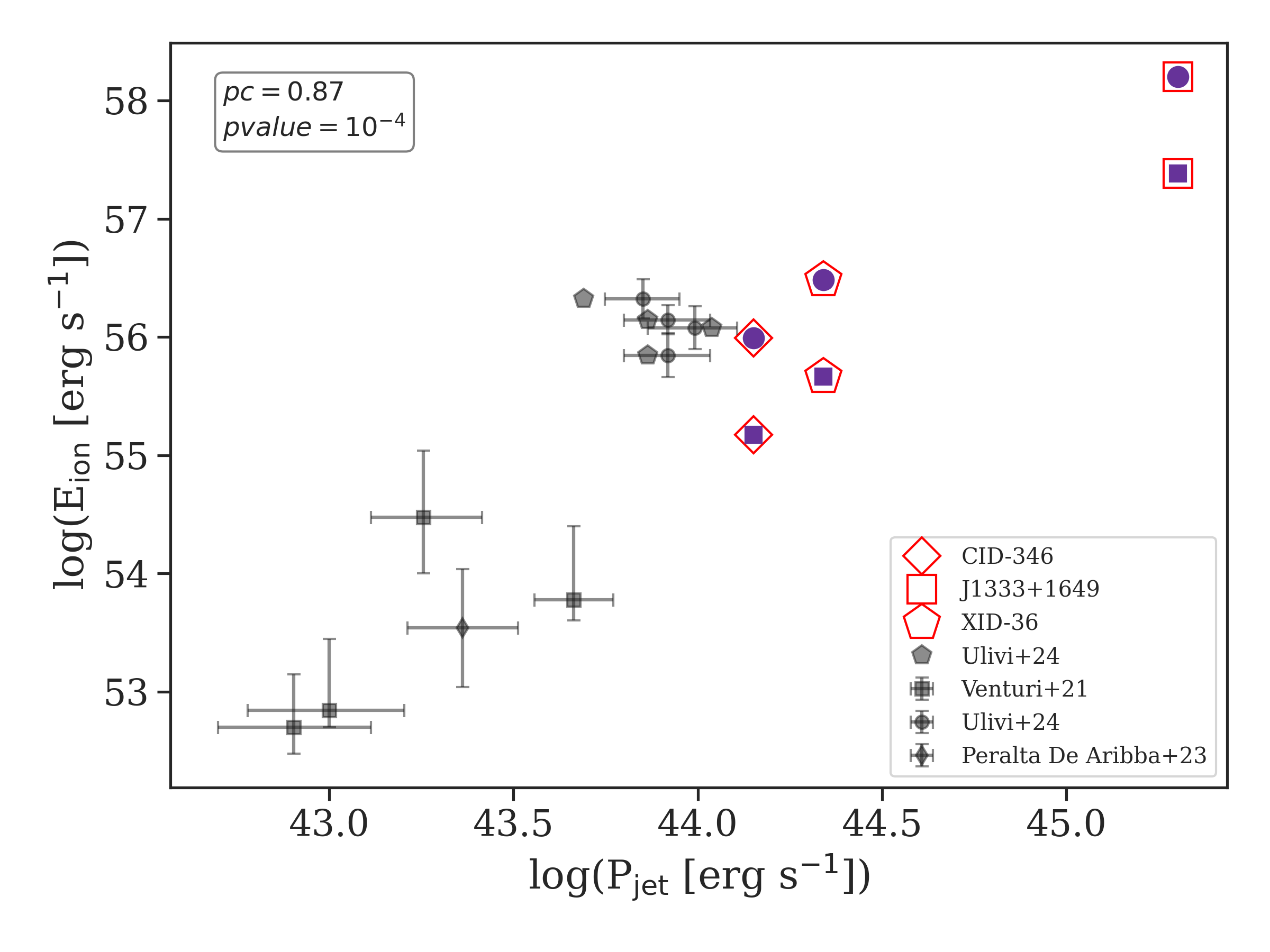}
\\
	\includegraphics[width=0.5\textwidth]{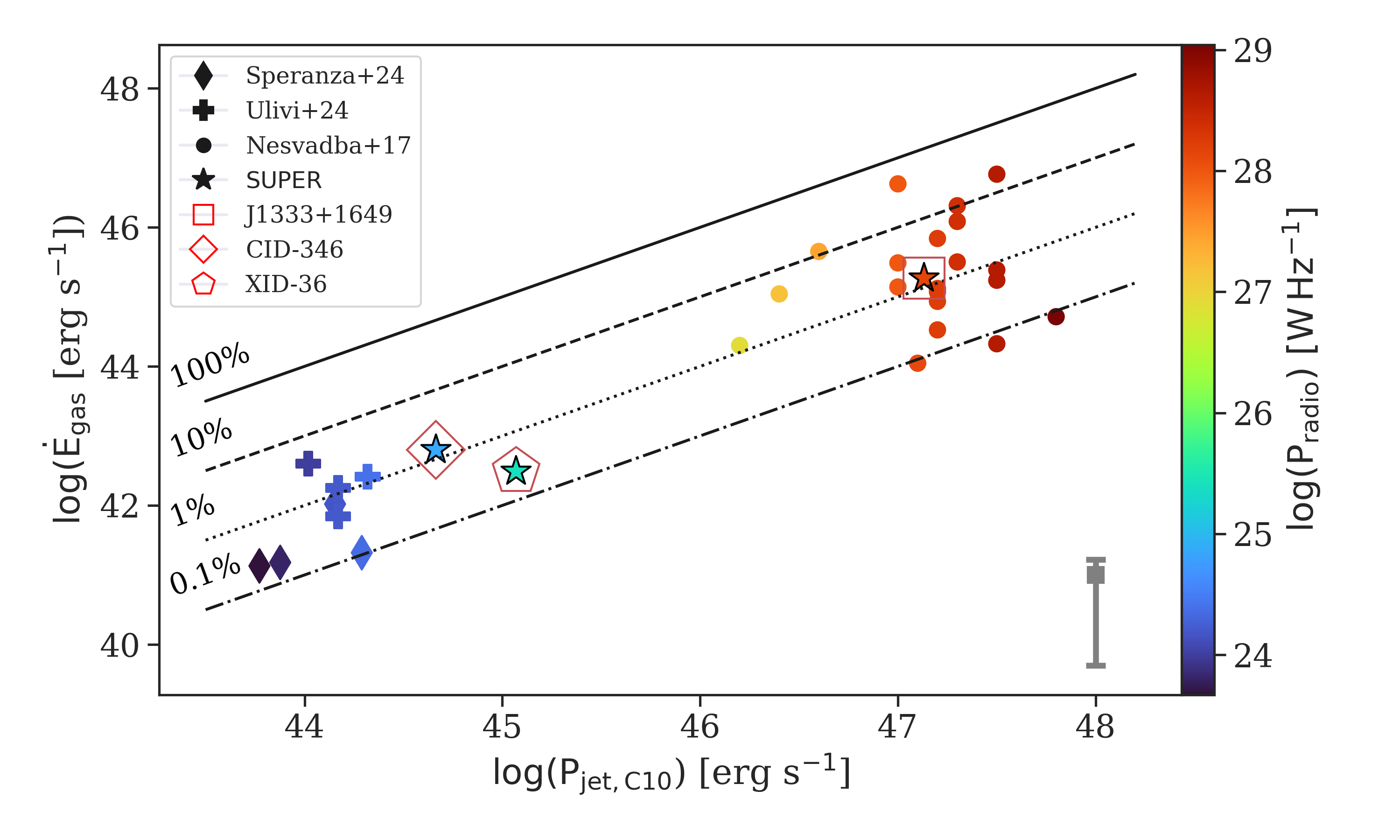}

    \caption{ Kinetic energy of the ionized gas versus radio jet 
 power. Top: Kinetic energy of the ionized gas in the disturbed region as a function of jet kinetic power (assuming the radio emission is attributed to jets) for our targets,  \citet{Ulivi+24} sample and their data compilation \citep[i.e.,][]{Venturi+21,Peralta+23}. The circles for our sample represent the kinetic energy values obtained considering W$_{80}$ and the squares represent those obtained considering the velocity dispersion. Bottom: Kinetic energy injection rates as a function of inferred radio jet  power (assuming the radio emission is attributed to jets) for our  targets, \citet{Ulivi+24}, \citet{Speranza+24} and \citet{Nesvadba+17}. The values for our sample were calculated assuming a gas density of 500 cm$^{-3}$, the same used by \citet{Nesvadba+17}. The gray bar indicates the minimum and maximum variation in the outflow kinetic power values, considering electron densities of 10$^{4}$ cm$^{-3}$ and 300 cm$^{-3}$, respectively. The color bar shows the 1.4 GHz radio power. The diagonal lines  indicate energy coupling efficiencies between the radio jet and the gas: 100\%, 10\%, 1\%, and 0.1\%.}
    
    \label{fig:ion_jet}
\end{figure}

We aim to explore how our sample compares to previous studies regarding the relative energetics of the outflows and the associated radio power. Specifically, we evaluate whether the observed outflows could be energetically driven by radio jets, under the assumption that the radio emission originates from jet activity. We perform this comparison with the  samples of  \citet{Venturi+21}, \citet{Ulivi+24}, \citet{Peralta+23}, \citet{Nesvadba+17} and \citet{Speranza+24}. These studies investigated systems that host low-power AGN radio jets (see Figure \ref{fig:ion_jet}) at low redshift ($z<0.15$), which exhibit significantly enhanced gas velocity dispersions on kiloparsec scales in directions perpendicular to both the jet axis and the ionization cone. Their main conclusion is that this phenomenon could result from the interaction between the radio jets and the ISM in the galaxy disk. This interpretation is also supported by simulations showing that low-power jets with small inclinations relative to the galactic disk can significantly disturb the ISM, with the strongest impact occurring perpendicular to the jet propagation axis \citep{Mukherjee+18,Talbot+22,Meenakshi+22}.  Meanwhile, \citet{Speranza+24} analyzed quasars and reported that, in sources with extended radio emission, the radio structures are well aligned with the outflows. This alignment suggests that low-power jets may play a role in compressing and accelerating the ionized gas in these radio-quiet quasars.

\citet{Ulivi+24} investigated whether such jet-driven perturbation of the ISM is feasible by comparing the kinetic energy of the ionized gas in the disturbed region with the observed jet kinetic power. By combining these measurements with those of \citet{Venturi+21}, they identified a correlation indicating that the kinetic energy of the ionized gas is closely linked to the jet kinetic power, supporting the scenario in which the jet is the mechanism behind the observed increase in gas velocity dispersion. Following \citet{Ulivi+24}, we also estimated the jet kinetic power using the equation obtained from \citet{Birzan+08}. We compared it with the kinetic energy of the ionized gas, in order to investigate if the radio jet could inject sufficient energy into the ISM to drive the observed outflows in our  sample. The kinetic energy of the ionized gas was determined using \(\rm E_{\text{ion}} = \frac{M_{\text{ion}} \, \sigma_{\text{ion}}^2}{2} \), where the ionized gas mass ($\rm M_{ion}$) is the same outflow mass ($\rm M_{out}$) determined by \citet{Kakkad+20} and \citet{Tozzi+24} considering a gas density of \( 500\,\text{cm}^{-3} \) and solar metallicity; $\sigma$ is the velocity  dispersion  of the  ionized gas. We also estimated the kinetic energy of the disturbed gas using \(\rm E_{\text{ion}} = \frac{M_{\text{ion}} \, W_{\text{80}}^2}{2} \).

In Figure \ref{fig:ion_jet}, we present the values obtained for our sample and those reported in the literature. By combining these samples, we observe a correlation between the kinetic energy of the ionized gas and the jet kinetic power, consistent with the findings of \citet{Ulivi+24}. We found a Pearson correlation coefficient of 0.93 and a p-value of $10^{-5}$ combining the values of our sample with those of the literature, when we use \weighty\ to calculate the kinetic energy of our sample. For the kinetic energy estimated through the velocity dispersion, we found a correlation coefficient of 0.87 and p-value  of  $10^{-4}$. These results suggest that, in our sources as well, the radio jet may be responsible for the disturbed kinematics of the gas, as evidenced by the high \weighty\ values, which in our case are linked to outflows. These findings support the idea that radio jets may act as the driving mechanism behind the observed outflows.

\citet{Nesvadba+17} compared the  kinetic energy injection rates with kinetic jet power for a sample of high radio power quasars, they found that most objects exhibit coupling efficiencies around 1\%, meaning that approximately 1\% of the jet kinetic power is transferred to the surrounding gas in the form of outflow kinetic energy, although the values show considerable scatter. Furthermore, the authors report that no galaxy exhibits a coupling efficiency greater than 100\%, indicating that no additional energy source beyond the radio jet is required to account for the observed kinetic energy in the gas.  Using the calibration of \citet{Cavagnolo+10} to estimate the kinetic power of the radio jets, the same adopted by \citet{Nesvadba+17}, we present in Figure \ref{fig:ion_jet} the radio jet kinetic powers, the ionized gas  kinetic energy injection rates, and the 1.4 GHz radio powers for  \citet{Nesvadba+17},  \citet{Ulivi+24} and our  own sample.  Our objects exhibit a coupling efficiency between the radio jet power and the kinetic energy of the ionized gas that is comparable to that of the galaxies in the  \citet{Ulivi+24} sample, lying around  $\sim$1\%. This value is consistent with the typical range reported by \citet{Nesvadba+17},  where most galaxies show coupling efficiencies close to this values, though with considerable scatter. 

Combining the results of the total energy of the ionized gas (Figure \ref{fig:ion_jet}) with the estimates of coupling efficiency, we find that although the overall coupling between the jet and the ionized gas is relatively low, it might still be sufficient to drive the observed outflows. However,  there are important caveats to consider when interpreting these results.  Notably, both the radio jet power and the outflow energetics carry significant uncertainties \citep[e.g.,][]{Harrison+18,Ward+24}.  Nevertheless, the observed systems follow similar trends to those seen in lower-redshift AGN and high radio power AGN at cosmic noon. Assuming a typical coupling efficiency of $\sim$1\%, the inferred jet power appears sufficient to drive the observed outflows, though we are only probing a single gas phase in this study. 

\begin{figure}    	\includegraphics[width=0.5\textwidth]{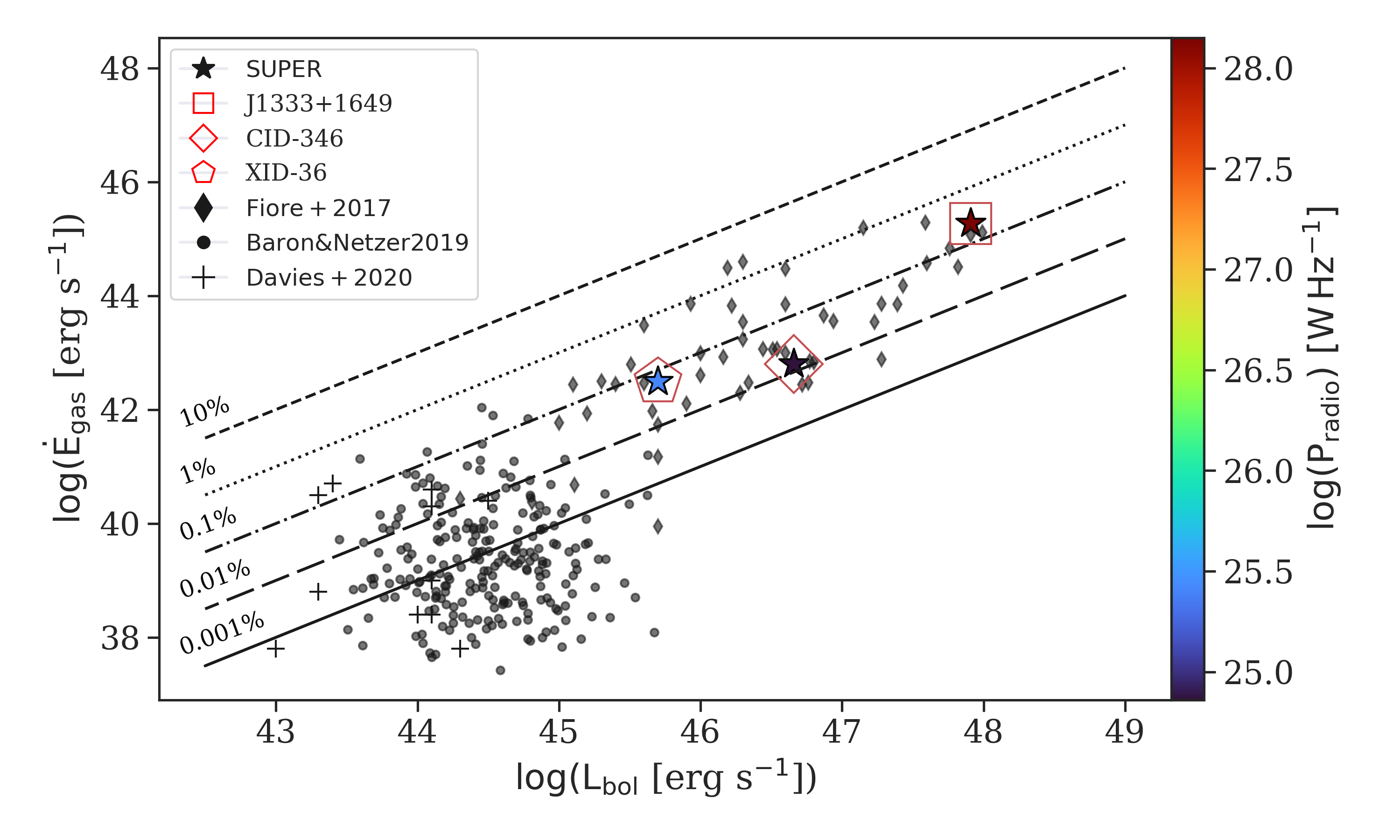}
    \caption{Kinetic energy injection rates of the ionized gas versus AGN bolometric luminosity for our  targets together with comparison samples from the literature, such as \cite{Fiore+17}, \citet{Baron+19} and \citet{Davies+20}. The color bar shows the 1.4 GHz radio power. The diagonal lines  indicate energy efficiencies: 10\%, 1\%, 0.1\%, 0.01\%, and 0.001\%.}
    
    \label{fig:ion_lbol}
\end{figure}

Since these quasar exhibit high bolometric luminosities, and \citet{Kakkad+20} concluded for two of our objects that the outflows could be driven by radiatively winds, it is important to test this scenario. In Figure \ref{fig:ion_lbol}, we therefore compare the kinetic power of the outflows ($\dot{E}_{\rm gas}$) with the bolometric luminosity derived from SED fitting ($L_{\rm bol}$), together with coupling efficiency lines. Our targets, as well as comparison samples from the literature, lie in the range $\dot{E}_{\rm gas}/L_{\rm bol} \sim 0.01\% - 1\%$, which is consistent with expectations for radiatively driven winds \citep[e.g.,][]{Fiore+17,Baron+19,Davies+20}. This suggests, based purely on these energetic arguments, that radiative driving provides a plausible mechanism for powering the observed outflows in these quasars, in addition to the possibility of jet-driven.

\section{Conclusions}
\label{sec:conclusions}
We have analyzed new 6.2 GHz VLA radio observations for three typical radiative AGNs with a radio power at 1.4 GHz of $\sim 10^{24.8}-10^{28.2}$ W Hz$^{-1}$ and combined them with \oiii\ data from SINFONI previously analyzed by our group \citep[see][]{Kakkad+20,Vietri+20,Lamperti+21,Tozzi+24}. These data enabled us to explore the radio ionized gas connection on a few kiloparsec scales. Moreover, we also used radio data from the literature of VIPS, MOJAVE, and VLA-COSMOS. Our primary goal was to conduct a pilot study to investigate the connection between radio emission and ionized gas at cosmic noon in a sample of three quasars that exhibit previously identified ionized gas outflows. The key findings presented in this paper are summarized as follows:

\begin{itemize}
\item Among our targets, two exhibit extended radio structures at a significance level of at least 36$\sigma$ for J1333+1649 (Figure \ref{fig:J1333_radio}) and 4$\sigma$ for CID-346 (Figure \ref{fig:cid346_radio}) in the VLA A-array data. In the case of J1333+1649, the extended structure (seen at $\sim$0.5\arcsec or $\sim4.16$ kpc ) aligns with the smaller-scale collimated jet ($\sim$0.01\arcsec-0.02\arcsec or $\sim0.08-0.17$ kpc) detected in VIPS and MOJAVE observations, suggesting a jet propagating across different spatial scales. XID-36 is the only target that exhibits a compact radio structure (Figure \ref{fig:XID36_radio}).

\item We find a connection between radio emission and ionized gas in our VLA A-array and  \oiii\ spatially resolved maps, with both components overlapping and exhibiting the same spatial extent for all quasars in our sample (Figure \ref{fig:J1333_oiii_radio}). Furthermore, the position angles of the radio emission and ionized gas exhibit offsets of less than 30$^{\circ}$ for the two galaxies where extended radio structure  is detected. Since the kinematics of \oiii\ in all three objects is dominated by outflows, we also observed that radio emission is directly associated with the outflows.

\item Our results indicate a correlation between the radio PA and the ionized gas PA at cosmic noon, as previously observed for moderate radio power quasars at low redshift and high radio power sources at cosmic noon. By combining our sample with low redshift sources (Figure \ref{fig:pa_jarvis}), we confirmed this correlation through the Pearson test, further supporting the connection between radio emission and ionized gas and suggesting that this could also be a common phenomenon at cosmic noon.

\item Jet kinetic power correlates with the kinetic energy of the ionized gas in our sample (Figure \ref{fig:ion_jet}), consistent with previous studies. The estimated coupling efficiency is around 1\%, similar to values reported in the literature \citep[e.g.,][]{Nesvadba+17}, suggesting that even modest coupling may be sufficient to drive the observed outflows. This supports the scenario of jet-driven outflows, though radiative driving remains a viable alternative in some cases.
\end{itemize}

Our spatially resolved results provide evidence of a connection between radio emission and ionized gas outflows, likely driven by the interaction between the radio jet and the ISM, although shocks from outflows launched by radiatively driven winds cannot be ruled out in all cases. This interaction appears to induce outflows in intermediate to high radio power quasars at cosmic noon. Our findings explored a new redshift and radio power regime for the connection between ionized gas and radio emission. This result reinforces the importance of considering jet-driven feedback in regulating gas dynamics across different cosmic epochs for radiative AGNs with moderate radio powers. Thus, our findings complement previous results in the literature across both redshift and radio power regimes. Future works should aim to explore large samples of typical sources at cosmic noon by combining high-resolution radio imaging with spatially resolved multi-phase outflow tracers to assess how common such jet-driven feedback processes are in shaping galaxy evolution.

\section*{Data availability}
The VLA radio images underlying this article are available at \href{https://doi.org/10.25405/data.ncl.30344950}{https://doi.org/10.25405/data.ncl.30344950}. 

\begin{acknowledgements}
      GSI acknowledges financial support from the Fundação de Amparo à Pesquisa do Estado de São Paulo (FAPESP), under Projects 2022/11799-9 and 2024/02487-9. AN, CMH and VAF acknowledge funding from an United Kingdom Research and Innovation grant (code: MR/V022830/1). EB and GC acknowledge financial support from INAF under the Large Grant 2022 ``The metal circle: a new sharp view of the baryon cycle up to Cosmic Dawn with the latest generation IFU facilities’' and the GO grant ``A JWST/MIRI MIRACLE: Mid-IR Activity of Circumnuclear Line Emission’’. EB acknowledges INAF funding through the “Ricerca Fondamentale 2024” program (mini-grant 1.05.24.07.01). MP acknowledges support through the grants PID2021-127718NB-I00, PID2024-159902NA-I00, and RYC2023-044853-I, funded by the Spain Ministry of Science and Innovation/State Agency of Research MCIN/AEI/10.13039/501100011033 and El Fondo Social Europeo Plus FSE+. We thank Dan Jackson-Thomas for sharing his preliminary investigation of some of the data from this study and Sean Dougherty for proposing for the observations. IL acknowledges support from PRIN-MUR project ``PROMETEUS''  financed by the European Union -  Next Generation EU, Mission 4 Component 1 CUP B53D23004750006. We thank Paolo Padovani for comments. We would also like to thank the referee for their valuable comments and suggestions, which helped improve this work.
\end{acknowledgements}

\bibliographystyle{aa} 
\bibliography{lib.bib}

\onecolumn 
\begin{appendix} 

\section{Sample properties}

\begin{table}[H]
\centering	
\caption{Sample properties obtained from \citet{Circosta+18}.}

\begin{tabular}{ccccccccc}
\hline
\rule{0pt}{3ex} 
Name & RA[J2000] & Dec[J2000] & $z_{spec}$ & log $\frac{L_{FIR}}{erg s^{-1}}$ & log $\frac{L_{bol}}{erg s^{-1}}$ & log $\frac{L_{[2-10 keV]}}{erg s^{-1}}$ & log $\frac{P_{1.4GHz}}{W Hz^{-1}}$ & q$_{24obs}$ \\
\\
(1) & (2) & (3) & (4) & (5) & (6) & (7) & (8) & (9) \\
\\
\hline
\\
J1333+1649 & 13:33:35.79 & +16:49:03.96 & 2.089 & - & 47.91$\pm$0.02 & 45.81$^{+0.07} _{-0.06}$ & 28.15$\pm$0.01 & -2.06 \\
\\
CID-346 & 09:59:43.41 & +02:07:07.44 & 2.219 & 46.13$\pm$0.06 & 46.66$\pm$0.02 & 44.47$^{+0.08}_{-0.09}$ & 24.86$\pm$0.07 & 0.05 \\
\\
XID-36 & 03:31:50.77 & -27:47:03.41 & 2.259 & 45.84$\pm$0.02 & 45.70$\pm$0.06 & 43.84$^{+0.31}_{-0.63}$ & 25.40$\pm$0.01 & -1.00 \\
\\
\hline
\end{tabular}
    \label{table:sample}
    \tablefoot{(1) Target name; (2) Right ascension; (3) Declination; (4) Redshift; (5) FIR luminosity in the 8-1000 $\mu$m range; (6) AGN bolometric luminosity; (7) Absorption-corrected X-ray luminosity in the hard band (2-10 keV); (8) Radio power at 1.4 GHz; (9) q$_{24 obs}$ = log(S$_{24\mu m}$/S$_{r}$), where S$_{24\mu\text{m}}$ and S$_{r}$ are the observed 24 $\mu\text{m}$ and 1.4 GHz flux densities.}
    \end{table}

\section{Specifications of VLA observations}

In this section, we present the observation dates along with their properties, including integration time and phase calibrator, for each object in the sample and for each VLA configuration (A and B-array). These details are summarized in Table \ref{tab:obs}. We also  present the images  properties for each target and array configuration (Table \ref{table:observations})
\\
\begin{table}[H]

\caption{Observation dates of the sources with the VLA in A and B configurations.}  
\centering	
\begin{tabular}{c c c c c c c} 
\hline \rule{0pt}{3ex} 
Name & \multicolumn{2}{c}{Obs. Dates} & \multicolumn{2}{c}{Integration time (min)} & \multicolumn{2}{c}{Phase calibrators} \\ 
& A-array  & B-array & A-array & B-array & A-array & B-array \\ 
(1) & (2) & (3) & (4) & (5) & (6) & (7) \\ \\ \hline \\ 
J1333+1649 & 2023-06-29 & 2023-01-13 & 22 & 15 & J1415+1320  & J1333+1649 \\  
CID-346 & 2023-06-29 & 2023-01-13 \& 2023-01-21 &  154 & 104 & J0943-0819 & J0943-0819 \\  
XID-36 & 2023-07-01 \& 2023-07-03 & 2023-01-15 \& 2023-01-17 &44  &30  & J0240-2309 & J0348-2749 \\  
\\ \hline
    \end{tabular}
    \label{tab:obs}
    \tablefoot{ Column (1) lists the source names; Column (2)-(3) provides the observation dates for the A-array and B-array configuration; (4)-(5) The total on-target integration times (in minutes) for each target for A-array and B-array configuration; (6)-(7) Phase calibrators for each target for A-array and B-array.}
    \end{table}

\begin{table}[H]
\caption{Summary of radio images parameters.}
\centering	
\begin{tabular}{ccccccccccc}
\hline
\rule{0pt}{3ex}  
Name  & Array & Beam  Size & Beam PA  & Robust &  RMS  & Image size & Pixel scale \\
&  &  & ($^{\circ}$) &  weighting  & ($ \mu\text{Jy}/\text{beam} $) & (arcsec) & (mas)\\ 
\\
(1) & (2) & (3) & (4) & (5) & (6) & (7) & (8)\\  
\\
\hline 
\\
J1333+1649 & A & 0.35\arcsec $\times$ 0.25\arcsec & -66.4 & 0 & 18.7 & 125\arcsec $\times$125\arcsec & 50 \\  
 & B &  0.90\arcsec $\times$ 0.73\arcsec & 3.7 & 0 &  77.0 & 260\arcsec $\times$260\arcsec &  130 \\
CID-346 & A &  0.39\arcsec $\times$ 0.34\arcsec & 14.2 & +1 & 2.0 & 125\arcsec $\times$125\arcsec & 50 \\  
  & B &  0.98\arcsec $\times$ 0.90\arcsec & -19.4  & 0 & 3.5 & 425\arcsec $\times$425\arcsec & 170 \\  
XID-36 & A & 0.67\arcsec $\times$ 0.30\arcsec  & -6.3  & +1 & 4.5  & 125\arcsec $\times$125\arcsec & 50 \\
 & B   & 1.95\arcsec $\times$ 0.68\arcsec  & 8.3 & 0 & 7.0 &  425\arcsec $\times$425\arcsec & 170 \\  
\\
\hline  
\end{tabular}
    \label{table:observations}
        \tablefoot{  Columns: (1) AGN identification; (2) Array configuration; (3) Beam size; (4) Beam position angle; (6) Robust weighting parameter; (7) Image region; and (8) Image pixel scale.}
        \end{table}

\section{Radio images from the VLA A and B-Array on a large spatial scale}
\label{Appendix_radio_images}
\begin{figure*}[htbp]
  	\includegraphics[width=\textwidth]{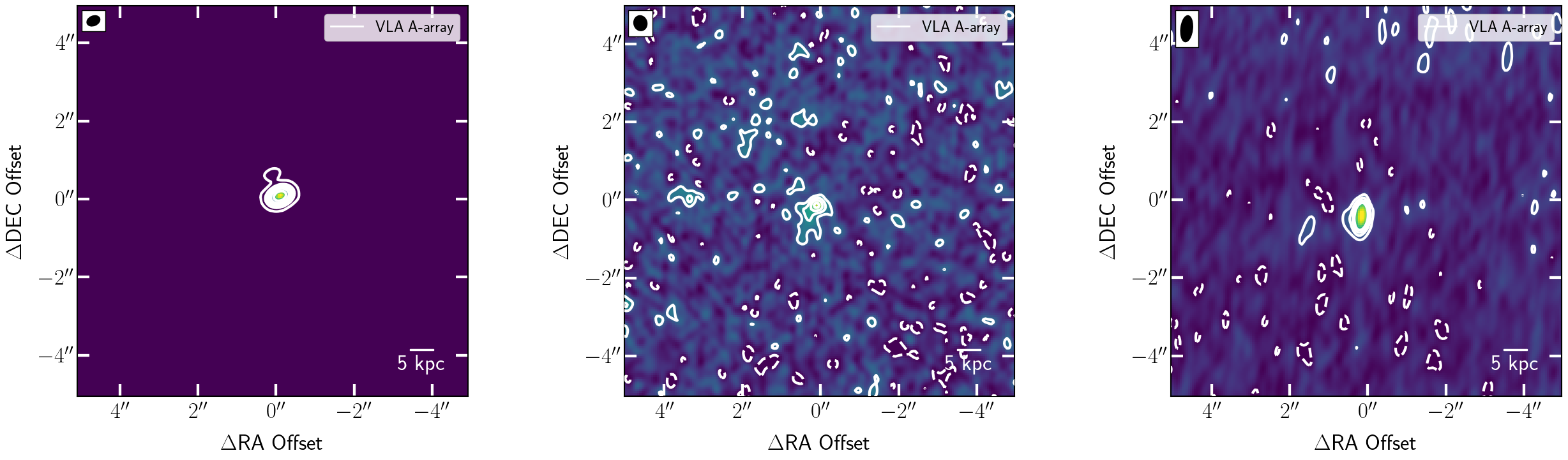}
    \\
      	\includegraphics[width=\textwidth]{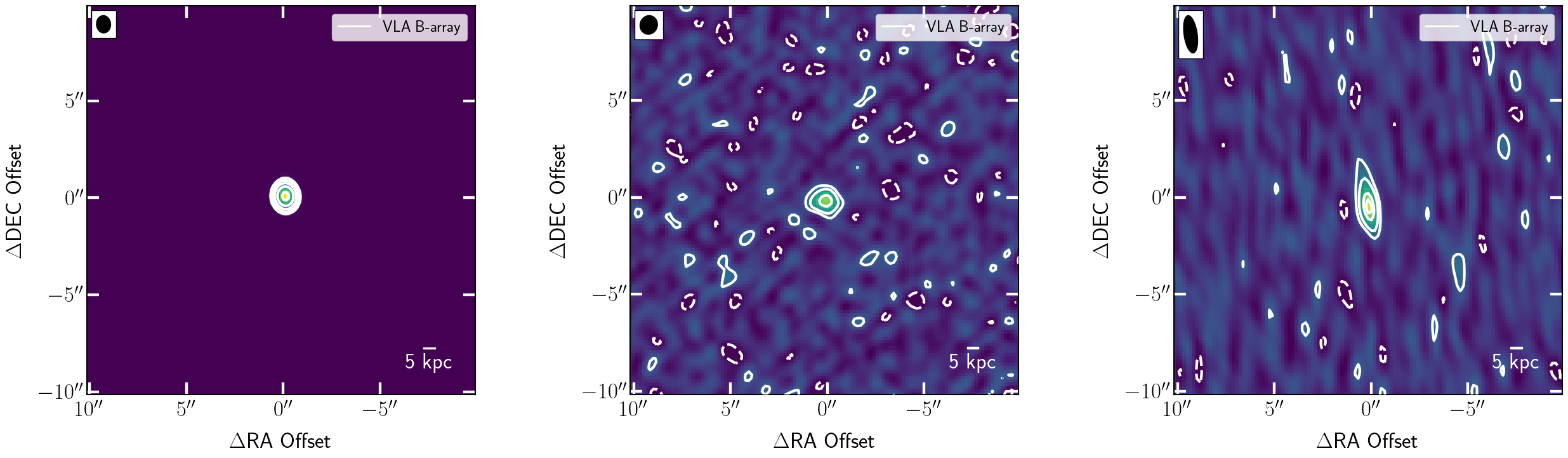}
   \caption{Radio contours maps of images for VLA A-array (top)  and VLA B-array (bottom) with  a larger field of view. The beam is represented by a ellipse at each image. Left: VLA A-array (top) and B-array (bottom) maps for J1333+164 with contours plotted at [36, 512, 1024, 2048, 4096, 8192]$\sigma$ and [128, 256, 512,1024, 2048]$\sigma$, respectively. Middle: VLA A-array (top) and B-array (bottom) maps for  CID-346 with contours plotted at [-2, 2, 4, 6, 8]$\sigma$ and [-2, 2, 4, 8]$\sigma$, respectively. Right: VLA A-array (top) and B-array (bottom) maps for  XID-36 with contours plotted at [-2, 2, 4, 6, 8]$\sigma$ and [-2, 2, 4, 8, 10]$\sigma$, respectively. }
    
    \label{fig:large_vla}
\end{figure*}

To assess the presence of large-scale radio structures associated with the central source in the VLA data, we generated A  and B-array images (Fig. \ref{fig:large_vla}) for the three objects in our sample, adopting a larger field of view than that shown in Figures  \ref{fig:J1333_radio}-\ref{fig:XID36_radio}. These images do not reveal any significant radio emission attributable to the central source on larger scales in either configuration.

\section{Position angles measured for the radio and ionized gas data}

\begin{table}[H]
\caption{Position angle for the extended radio structures and the ionized gas (morphological and kinematical). }
\centering
\begin{tabular}{lcc}
\hline
Image & PA J1333+1649 ($^{\circ}$) & PA CID-346  ($^{\circ}$) \\
\hline
VLA-A array             & 22 $\pm$ 20 &  130 $\pm$ 17 \\
VIPS                    & 19 $\pm$ 6  & -- \\
MOJAVE                  & 18 $\pm$ 5  & -- \\
Ionized gas             & 49 $\pm$ 22 & 130 $\pm$ 30 \\
Kinematic & 2 $\pm$ 26 & 158 $\pm$ 9.0\\
VLA-COSMOS               & --           & 148 $\pm$ 6 \\
\hline
\end{tabular}

\label{table:pa_combined}
        \tablefoot{The kinematic PA was measured using the PaFit package from \citet{Krajnovic+06}. A dash (--) indicates that the measurement is not available for that object.  The position angles are measured relative to the north direction of the image, anticlockwise. XID-36 does not show any extended structures. }
        \end{table}

\section{SINFONI H-band continuum maps}
\newpage
\begin{figure*}

    	\includegraphics[width=0.49\textwidth]{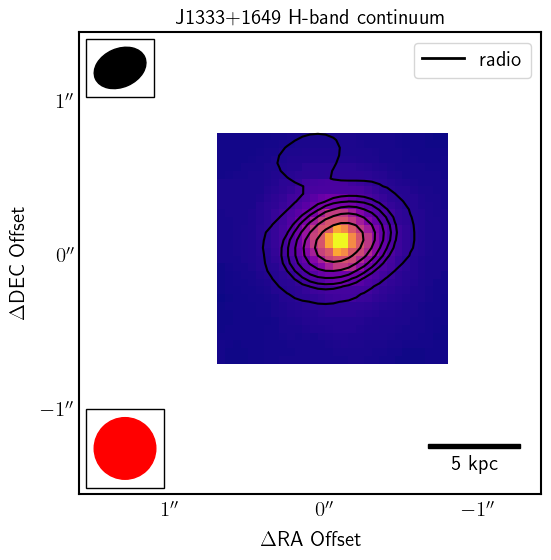}
	\includegraphics[width=0.49\textwidth]{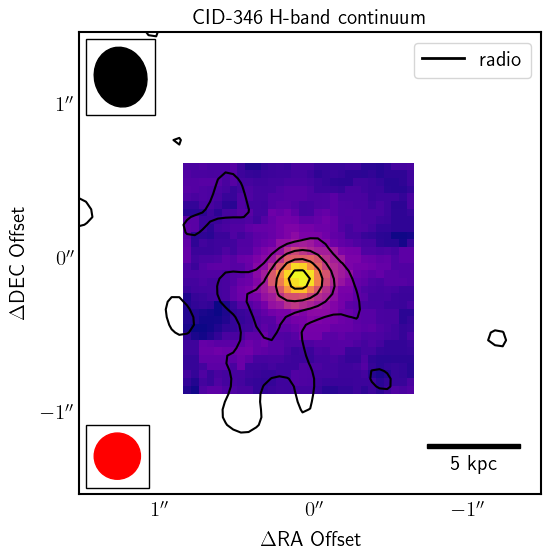}

    \caption{SINFONI H-band continuum maps for J1333+1649 (left panel)  and CID-346 (right panel). The contours plotted at [36, 512, 1024, 2048, 4096, 8192]$\sigma$ and [2, 4, 6, 8]$\sigma$ show the VLA-A array radio data for J1333+1649 and CID-346, respectively. The PSF and beam for the ionized gas maps and radio image are represented by a red and a black ellipse, respectively. For XID-36, the H-band continuum is not detected.}
    \label{cont_map}
\end{figure*}

\section{Molecular gas observations for CID-346}
\label{molecular_gas}
The CID-346 observations were obtained with the Medium Resolution Spectrometer (MRS) of the Mid-Infrared Instrument (MIRI) of JWST, as part of the Cycle 1 GO program 2177 (PI: Mainieri). $\rm H_{2}$ $\rm2.12~\mu\mathrm{m}$ observations for CID-346 were carried out on 20 November 2023, covering the wavelength range $\rm 6.53-7.65~\mu\mathrm{m}$, which corresponds to  $\rm2.02-2.37~\mu\mathrm{m}$ in the rest frame. The CID-346 data reduction and analysis were performed by \citet{Kakkad+25}, who reported the first detection of hot molecular gas in CID-346 via the $\rm H_{2}$ $\rm2.12~\mu\mathrm{m}$. In Section \ref{sec:Spatial_alignment}, we contextualize the results of \citet{Kakkad+25} for CID-346 regarding the molecular gas with our findings for the ionized gas and radio emission.

\begin{figure*}[h]
    \centering
    	\includegraphics[width=0.5\textwidth]{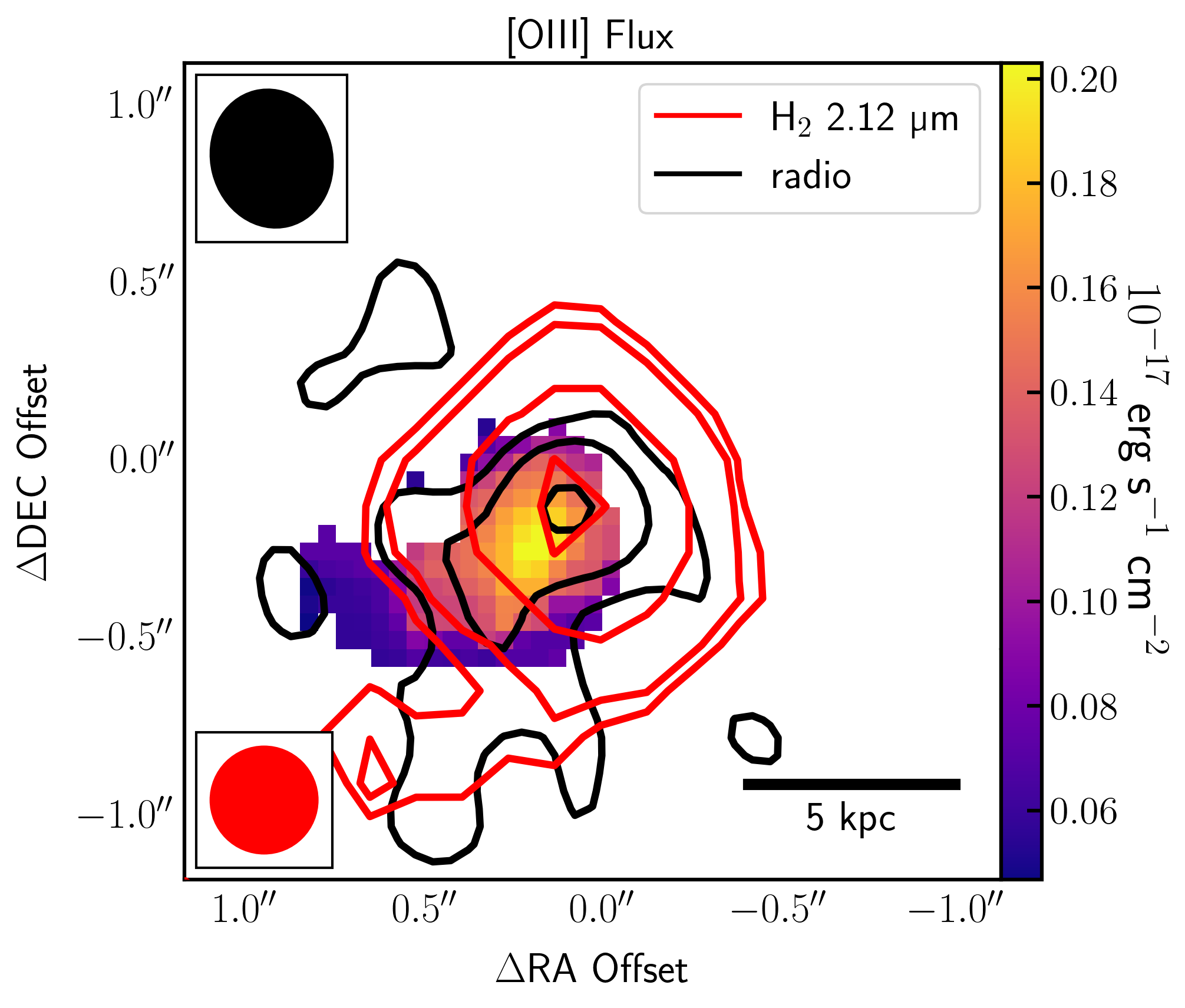}

    \caption{\oiii\  map for CID-346.  The PSF and beam for the ionized gas maps and radio image are represented by a red and a black ellipse, respectively. The black contours at [2, 4, 8]$\sigma$ show the VLA-A array radio data, while the red contours show the  $\rm H_{2}$ $\rm2.12~\mu\mathrm{m}$ emission obtained from \citet{Kakkad+25}.}
    \label{h2_map}
\end{figure*}

\end{appendix}

\end{document}